\documentstyle[aps]{revtex}
\tightenlines
\newtheorem{lemma}{Lemma}
\newtheorem{theorem}{Theorem}
\newtheorem{definition}{Definition}
\begin{document}
\draft
\title{What is Possible Without Disturbing Partially Known Quantum States?}
\author{Masato Koashi and Nobuyuki Imoto}
\address{CREST Research Team for Interacting Carrier Electronics,
School of
Advanced Sciences, \\
~The Graduate University for Advanced Studies (SOKEN),
Hayama, Kanagawa, 240-0193, Japan}
\maketitle
\begin{abstract}
Consider a situation in which a quantum system is secretly 
prepared in a state chosen from the known set of 
states. We present a principle that gives a definite distinction
between the operations that preserve the states of the system and 
those that disturb the states. The principle is derived by 
alternately applying a fundamental property of classical signals
and a fundamental property of quantum ones. 
The principle can be cast into a simple form by using a decomposition 
of the relevant Hilbert space, which is uniquely determined 
by the set of possible states. The decomposition implies 
the classification of the degrees of freedom of the system 
into three parts depending on how they store the information 
on the initially chosen state: one storing it classically,
one storing it nonclassically, and the other one storing no
information. Then the principle states that the nonclassical 
part is inaccessible and the classical part is read-only if we are
to preserve the state of the system.
From this principle,
many types of no-cloning, no-broadcasting, and no-imprinting conditions
can easily be derived in general forms including mixed states.
It also gives a  unified view on how various schemes of quantum
cryptography work. The principle helps to derive optimum amount of
resources (bits, qubits, and ebits) required in data compression 
or in quantum teleportation of mixed-state ensembles.

\end{abstract}
 

\section{Introduction}

Quantum mechanics pose fundamental restrictions when one reads out
information from a quantum system. The most basic rule is well known 
---if one reads out information from a quantum system in an unknown
initial state, the quantum state of the system will
change\cite{wootters82}. 
Recent development of quantum information theory
proposes various schemes of handling information through quantum systems,
and understanding of more detailed rules seems to become an important
issue. One particular direction  of such investigation is the cases when
the initial state is partially 
known\cite{yuen86,bennett92feb,barnum96,koashi98nov}. 
In such situations, 
some operations can be done without introducing any disturbance on the 
original quantum system. One of the fundamental questions here is
the following: 
What kind
of information can be extracted, and what cannot be, without changing the
state? This problem is important in quantum cryptography, since the
initial state is chosen by the sender among a few definite states. The
problem is also directly related to the physical feasibility of cloning
(making a copy of the original) and imprinting (catching a trail without
affecting  the original) of partially known quantum states. So far, the
 conditions for the initial states allowing such tasks were
derived, such as broadcasting of mixed
states\cite{barnum96} and cloning of pure entangled 
states\cite{koashi98nov}. The proofs were based on the complicated 
series of inequalities related to the fidelity, and it is not always easy
to infer the conditions even for slightly different tasks.

In this paper, we present a principle that gives a definite distinction
between what one can do and what one cannot do without changing the
state of a system. Given a set of possible initial states, we 
propose a particular decomposition [Eq.~(\ref{decompositionha})]
of the system, which
classifies the degrees of freedom of the system into three parts,
according as how they hold the information on which one of the states is
chosen as the initial state.
The principle is then stated as the restriction to 
the access to each part. We provide a proof that clarifies the physical
origin of the principle---it is obtained by simply applying 
two fundamental theorems alternately, which respectively reflect the
basic property of classical signals (Theorem 1)
and that of quantum signals (Theorem 2). 
This principle can be applied to various problems of cloning
and imprinting of quantum states, and reveals conditions
for feasibility of various tasks such as no-imprinting
condition of mixed states. It also gives a good insight into the 
basic concepts of quantum  cryptography.

This paper is organized as follows. In Sec.~\ref{formulation},
we formulate the problem considered in this paper.
In Sec.~\ref{sec:basic}, we derive two theorems 
which reflect the basic property of classical signals 
and that of quantum signals. The latter one suggests 
a structure of Hilbert space in which tensor products and 
direct sums are involved, and we discuss notations to treat 
such structures in Sec.~\ref{sec:structure}. 
In Sec.~\ref{sec:main}, we repeatedly use the two basic theorems
and derive the main result, the property of the operations 
preserving a set of states. Sec.~\ref{sec:properties}
discusses properties of the decomposition used in stating the main 
theorem, such as its uniqueness and relation to well-known mathematical
concepts. 
In Sec.~\ref{sec:transfer}, the main theorem is restated in 
a scenario of faithful transfer,
which makes it convenient to apply the theorem 
to communication problems. 
In Sec.~\ref{sec:application}, we give  applications of the 
theorem to various problems of cloning, imprinting, 
quantum cryptography, quantum data compression, and teleportation.

\section{Formulation of the problem}
\label{formulation}

The main problem considered in this paper is described as follows.
Consider a quantum system A, which is described by a Hilbert space
${\cal H}_{\rm A}^\prime$.
Initially system A is secretly 
prepared in a state described by a 
normalized density operator $\rho_s$, one in the
known set of states 
$\{\rho_s\}_{s\in S}$. 
Here $S$ is the set of possible values of index $s$.
For example, if the initial state is chosen from 
$n$ states, $S$ is assumed be $\{1,2,\ldots,n\}$.
$S$ can also be a infinite set. We assume that 
$\{\rho_s\}_{s\in S}$ is supported by a subspace
with a finite dimension. This assumption is more precisely
stated as follows. Let us write the support of $\rho_s$ as 
${\rm Supp}(\rho_s)$, and define 
\begin{equation}
{\cal H}_{\rm A}\equiv \bigcup_{s\in S} {\rm Supp}(\rho_s).
\label{Supprhos}
\end{equation}
Then, the said assumption is that the dimension of
 ${\cal H}_{\rm A}$ be finite.

Next, we prepare an ancilla (an auxiliary system) E,
described by a Hilbert space
${\cal H}_{\rm E}$, 
in a standard quantum state 
$\Sigma_{\rm E}=|u\rangle_{\rm E}\langle u|$,  
and apply a unitary operation $U$ on 
${\cal H}_{\rm A}^\prime\otimes{\cal H}_{\rm E}$. 
After this operation, 
the marginal density operator
 of ${\cal H}_{\rm A}^\prime$ becomes 
\begin{equation}
{\cal T}_U(\rho_s)\equiv \mbox{Tr}_{\rm E}[U(\rho_s\otimes \Sigma_{\rm E} )U^\dagger].
\end{equation}  
What we seek is the requirement for $U$ to preserve 
the marginal density operator of A, namely,
${\cal T}_U(\rho_s)=\rho_s$ for all $s \in S$.
Note that what we concern here is not the whole property of 
$U:{\cal H}_{\rm A}^\prime\otimes{\cal H}_{\rm E}
\rightarrow {\cal H}_{\rm A}^\prime\otimes{\cal H}_{\rm E}$,
but that of the isometry given as its restriction, 
$U:{\cal H}_{\rm A}\otimes|u\rangle_{\rm E}
\rightarrow {\cal H}_{\rm A}^\prime\otimes{\cal H}_{\rm E}$.
Let ${\cal U}_{\rm all}$ be the set of all isometries from 
${\cal H}_{\rm A}\otimes|u\rangle_{\rm E}$
 to ${\cal H}_{\rm A}^\prime\otimes{\cal H}_{\rm E}$.
The problem here is thus to identify the subset 
\begin{equation}
{\cal U}_{\rm ND}\equiv\{U\in {\cal U}_{\rm all}|{\cal
T}_U(\rho_s)=\rho_s,
 {}^\forall s\in S\}.
\label{defund}
\end{equation}

It is convenient to construct a density operator  
 $\rho_{\rm all}$ 
from $\{\rho_s\}_{s\in S}$, satisfying the following conditions:
\begin{equation}
{\rm Tr}(\rho_{\rm all})=1,
\label{Trrhoall}
\end{equation}  
\begin{equation}
{\cal T}_U(\rho_{\rm all})=\rho_{\rm all} \;\; 
{}^\forall U\in {\cal U}_{\rm ND},
\label{NDrhoall}
\end{equation}
and 
\begin{equation}
{\rm Supp}(\rho_{\rm all})={\cal H}_{\rm A}.
\label{Supprhoall}
\end{equation}
When a probability distribution $p(s) (s \in S)$ over $S$
satisfying $p(s)>0$ for all $s\in S$ is assigned to the set 
$\{\rho_s\}_{s\in S}$, $\rho_{\rm all}$ can be constructed 
as an averaged state, namely, by
 a sum $\rho_{\rm all}=\sum_{s\in S} p(s) \rho_s$, or
by an integral $\rho_{\rm all}=\int ds p(s) \rho_s$.
Alternatively,
we can always pick up $n(\le \rm{dim}\; {\cal H}_{\rm A})$ 
states $\{\rho_1,\rho_2,\ldots,\rho_n\}$ from the set 
$\{\rho_s\}_{s\in S}$ such that 
${\rm Supp}(\sum_{i=1}^n\rho_i)={\cal H}_{\rm A}$. Then, 
setting $\rho_{\rm all}=\sum_{i=1}^n\rho_i/n$ satisfies
Eqs.~(\ref{Trrhoall})-(\ref{Supprhoall}).

\section{basic property of classical and quantum signals}
\label{sec:basic}
\subsection{Useful lemmas}

In this section, we introduce 
two lemmas, which will be frequently used in this paper.
\begin{lemma}
Let $O$ be an Hermitian
operator acting on ${\cal H}$, and $U$ be
a unitary operator on ${\cal H}\otimes {\cal H}_{\rm E}$
(or an isometry from ${\cal H}\otimes |u\rangle_{\rm E}$
to ${\cal H}\otimes {\cal H}_{\rm E}$) 
satisfying ${\cal T}_U(O)=O$. 
Then,
\begin{equation}
[P_+\otimes \bbox{1}_{\rm E}
,U](P_+ \otimes \Sigma_{\rm E} )=\bbox{0},
\label{lemma}
\end{equation}  
where $P_+$ is the projection onto the space spanned 
by the eigenvectors of $O$ with
positive eigenvalues.
\end{lemma}
This lemma implies that an operation preserving 
a Hermitian operator $O$ does not transfer the
eigenvectors of $O$ with positive eigenvalues to the space 
for nonpositive
eigenvalues. A proof is given as follows. 
Let us define $\bar{P}_+\equiv \bbox{1}-P_+$.
The operator $O$ can be 
decomposed
as $O=O_+ - O_-$ by a positive definite operator $O_+\equiv P_+ O$ and
a positive semidefinite operator $O_-\equiv-\bar{P}_+ O$. 
Since ${\cal T}_U$ is linear,
\begin{eqnarray}
&&{\rm Tr}[P_+{\cal T}_U(O)]=
{\rm Tr}[P_+{\cal T}_U(O_+)]
-{\rm Tr}[P_+{\cal T}_U(O_-)]
\nonumber \\
&&={\rm Tr}[{\cal T}_U(O_+)]-
{\rm Tr}[\bar{P}_+{\cal T}_U(O_+)]
-{\rm Tr}[P_+{\cal T}_U(O_-)].
\label{lem1pr1}
\end{eqnarray}
From ${\cal T}_U(O)=O$, we have 
\begin{equation}
{\rm Tr}[P_+{\cal T}_U(O)]={\rm Tr}[P_+O]={\rm Tr}[O_+].
\label{lem1pr2}
\end{equation}
On the other hand, 
since ${\cal T}_U$ is a trace-preserving map, we have
\begin{equation}
{\rm Tr}[{\cal T}_U(O_+)]={\rm Tr}[O_+].
\label{lem1pr3}
\end{equation}
Combining Eqs.~(\ref{lem1pr1})--(\ref{lem1pr3}), we obtain
\begin{equation}
{\rm Tr}[\bar{P}_+{\cal T}_U(O_+)]
+{\rm Tr}[P_+{\cal T}_U(O_-)]
=0.
\label{lem1pr4}
\end{equation}
Since ${\cal T}_U$ is a complete positive map, 
${\rm Tr}[\bar{P}_+{\cal T}_U(O_+)]\ge 0$ and 
${\rm Tr}[P_+{\cal T}_U(O_-)]\ge 0$. 
This means that
both terms in the lhs (left-hand side) of
Eq.~(\ref{lem1pr4}) are nonnegative, and we obtain 
${\rm Tr}[\bar{P}_+{\cal T}_U(O_+)]=0$.
This relation is also written as 
$\mbox{Tr}[QQ^\dagger]=0$ with 
$Q=(\bar{P}_+\otimes \bbox{1}_{\rm E})U(\sqrt{O_+}\otimes
 \Sigma_{\rm E} )$. This means $Q={\bf 0}$, or equivalently,
\begin{equation}
(\bar{P}_+\otimes \bbox{1}_{\rm E})U(P_+\otimes
 \Sigma_{\rm E} )
={\bf 0}.
\label{lem1pr5}
\end{equation}
Substituting $\bar{P}_+={\bf 1}-P_+$ completes the proof of Lemma 1.

\begin{lemma}
Let $\rho$ be a positive semidefinite operator
acting on ${\cal H}$.
Suppose that
its support  
${\rm Supp}(\rho)$ is written as a direct sum of 
two subspaces as 
${\rm Supp}(\rho)={\cal H}_1\oplus{\cal H}_2$,
 and let $P_i$ be
the projection  onto ${\cal H}_i(i=1,2)$. 
Let $U$ be
a unitary operator on ${\cal H}\otimes {\cal H}_{\rm E}$
(or an isometry from ${\cal H}\otimes |u\rangle_{\rm E}$
to ${\cal H}\otimes {\cal H}_{\rm E}$) 
satisfying ${\cal T}_U(\rho)=\rho$ 
and
$[P_1 \otimes \bbox{1}_{\rm E}, U] (P_1 
\otimes \Sigma_{\rm E})=\bbox{0}$.
Then
\begin{equation}
[P_2 \otimes \bbox{1}_{\rm E}, U] (P_2 
\otimes \Sigma_{\rm E})=\bbox{0}.
\label{lemma2}
\end{equation}
\end{lemma}
This lemma implies that if $U$ 
does not transfer the vectors in subspace ${\cal H}_1$ 
 to subspace ${\cal H}_2$,
$U$ does not include the transfer in the opposite way (${\cal H}_2$
 to ${\cal H}_1$).
Lemma 2 is proved as follows. 
The assumption 
$[P_1 \otimes \bbox{1}_{\rm E}, U] (P_1 
\otimes \Sigma_{\rm E})=\bbox{0}$ implies that
$(P_2 \otimes \bbox{1}_{\rm E})U (P_1 
\otimes \Sigma_{\rm E})=\bbox{0}$.
Using this,
we have 
\begin{eqnarray}
{\rm Tr}[P_2{\cal T}_U(\rho)]&=&
{\rm Tr}[P_2{\cal T}_U((P_1+P_2)\rho (P_1+P_2))]
\nonumber \\
&=&{\rm Tr}[P_2{\cal T}_U(P_2\rho P_2)].
\label{lem2pr1}
\end{eqnarray}
From ${\cal T}_U(\rho)=\rho$, we have 
\begin{equation}
{\rm Tr}[P_2{\cal T}_U(\rho)]
={\rm Tr}[P_2\rho].
\label{lem2pr2}
\end{equation}
Since ${\cal T}_U$ is a trace-preserving map, 
\begin{equation}
\mbox{Tr}[{\cal T}_U(P_2\rho P_2)]
={\rm Tr}[P_2\rho].
\label{lem2pr3}
\end{equation}
Combining Eqs.~(\ref{lem2pr1})--(\ref{lem2pr3}), we obtain
$\mbox{Tr}[{\cal T}_U(P_2\rho P_2)]=
{\rm Tr}[P_2{\cal T}_U(P_2\rho P_2)]$,
or equivalently, 
${\rm Tr}[\bar{P}_2{\cal T}_U(P_2\rho P_2)]={\bf 0}$
with $\bar{P}_2\equiv\bbox{1}-P_2$.
This relation is also written as 
$\mbox{Tr}[QQ^\dagger]=0$ with 
$Q=(\bar{P}_2\otimes \bbox{1}_{\rm E})U(\sqrt{P_2\rho P_2}\otimes
 \Sigma_{\rm E} )$. This means $Q={\bf 0}$, or equivalently,
\begin{equation}
(\bar{P}_2\otimes \bbox{1}_{\rm E})U(P_2\otimes
 \Sigma_{\rm E} )
={\bf 0}.
\label{lem2pr5}
\end{equation}
Substituting $\bar{P}_2={\bf 1}-P_2$ completes the proof of Lemma 2.

\subsection{Property of classical signals}

In this section, we derive a theorem which stems from 
a general property of classical signals. Before the derivation 
of the theorem, it is instructive to consider an example 
in the purely classical situation.
A classical counterpart of the problem considered here 
is obtained by replacing the requirement of 
preserving density operators to that of 
preserving probability distributions. 
Consider a purely
classical example, in which a signal $X$ is drawn from
either of the two probability distributions $p_1(x)$ and $p_2(x)$,
according to the value of $s(=1,2)$,
and a signal $\tilde{X}$ is then produced from the value 
of $X$ according to a rule that is independent of the value of $s$;
namely, if $X=x$, $\tilde{X}$ is set to 
$\tilde{X}=y$ with probability $p(y|x)$.
The probability distribution for $\tilde{X}$ is 
then given by $\tilde{p}_s(x)=\sum_{x^\prime} p(x|x^\prime)p_s(x^\prime)$.
 Let $K_0=\{x|p_1(x)+p_2(x)>0\}$ be the set of the possible
values of $X$. Let us divide $K_0$ into two sets,
$K_a\equiv\{x|p_1(x)>p_2(x)\}$ and 
$K_b\equiv\{x|p_2(x)\ge p_1(x), p_2(x)>0\}$.
A necessary condition for the transition matrix $p(y|x)$
in order that $\tilde{p}_s(x)$ coincides with $p_s(x)$
for either value of $s$ is that the transition must be made within 
each of the two sets $K_a$ and $K_b$, which is proved as follows.
 
Let us define $p^{(s)}(Z\in K)\equiv \sum_{x\in K} Prob\{Z=x\}
 (Z=X, \tilde{X}, K=K_a, K_b)$ 
as the probability that
the value of $Z$ belongs to $K$.
Consider quantities 
$d_a(X)\equiv p^{(1)}(X\in K_a)-p^{(2)}(X\in K_a)$
and
$p_b(X)\equiv p^{(1)}(X\in K_b)+p^{(2)}(X\in K_b)$,
and their changes in the transition $p(y|x)$, namely,
$\Delta d_a\equiv d_a(\tilde{X})-d_a(X)$
and 
$\Delta p_b\equiv p_b(\tilde{X})-p_b(X)$.
In order for $\tilde{p}_s(x)=p_s(x)$, these changes must be
zero. 
These changes are caused by the transition from 
$K_a$ to $K_b$ or vice versa, and $\Delta d_a$ is written as
the sum of two nonpositive parts,
\begin{equation}
\Delta d_a=-\sum_{y\in K_b} \sum_{x \in K_a}
p(y|x)[p_1(x)-p_2(x)]-\sum_{y\in K_a} \sum_{x \in K_b} 
p(y|x)[p_2(x)-p_1(x)].
\end{equation}
In order to satisfy $\Delta d_a=0$,
either part must be zero.
Since $p_1(x)-p_2(x)>0$ in the first part,
$p(y|x)$ with $y \in K_b$ and $x \in K_a$ must vanish. 
Under this
condition,
$\Delta p_b$ is contributed only by the transition from 
$K_b$ to $K_a$, and is given by 
\begin{equation}
\Delta p_b=-\sum_{y\in K_a} \sum_{x \in K_b} 
p(y|x)[p_1(x)+p_2(x)].
\end{equation}
Since  $p_1(x)+p_2(x)>0$,
$p(y|x)$ with $y \in K_a$ and $x \in K_b$ must also vanish 
in order to satisfy $\Delta p_b=0$.
Hence, preserving 
$p_1(x)$ and $p_2(x)$ requires that 
for any $y \in K_b$ and $x \in K_a$, $p(y|x)$ and  
$p(x|y)$ should vanish. The transition must be made within 
each of the two sets $K_a$ and $K_b$.

This argument almost directly applies to the quantum case, 
that is, we can show that  
any operation that preserves two different density operators,
$\rho$ and $\rho^\prime$, must act on two subspaces independently.
In order to represent this property in a simple form, 
we write the set of 
all isometries from 
${\cal H}\otimes|u\rangle_{\rm E}$
 to ${\cal H}\otimes{\cal H}_{\rm E}$ as ${\cal U}({\cal H})$,
where ${\cal H}$ is an arbitrary subspace. Then, 
the property is described by the following theorem.
\begin{theorem}
Let $\rho$ and $\rho^\prime$ be two density operators
for different states. Let ${\cal H}$ be the support of $\rho+\rho^\prime$,
and take the decomposition ${\cal H}={\cal H}_1\oplus {\cal H}_2$
where 
${\cal H}_1$ is the space spanned 
by the eigenvectors of
$O\equiv\rho/\mbox{Tr}(\rho)-\rho^\prime/\mbox{Tr}(\rho^\prime)$ with
positive eigenvalues.  Then, 
${\cal H}_1$ and ${\cal H}_2$ are nonzero subspaces, and 
any $U\in {\cal U}({\cal H})$ that satisfies 
${\cal T}_U(\rho)=\rho$ and 
${\cal T}_U(\rho^\prime)=\rho^\prime$
can be written as $U=U_1\oplus U_2$
with $U_i\in {\cal U}({\cal H}_i) (i=1,2)$.
\end{theorem}
For later convenience, the theorem allows for the possibility that 
$\rho$ and $\rho^\prime$ are unnormalized.
Theorem 1 is proved as follows. 
Since $\rho$ and $\rho^\prime$ represent different 
states, $O$ is nonzero. The form of $O$ implies that 
$O$ is a traceless Hermitian operator.
Hence $O$ has positive and negative eigenvalues, 
and ${\cal H}_1$ and ${\cal H}_2$ are nonzero spaces.
Next, suppose that 
$U\in {\cal U}({\cal H})$ satisfies 
${\cal T}_U(\rho)=\rho$ and 
${\cal T}_U(\rho^\prime)=\rho^\prime$.
Let $P_i$ be the projection onto ${\cal H}_i(i=1,2)$.
Since ${\cal T}_U$ is linear, ${\cal T}_U(O)=O$
and ${\cal T}_U(\rho+\rho^\prime)=\rho+\rho^\prime$.
From ${\cal T}_U(O)=O$, Lemma 1 leads to
\begin{equation}
[P_1\otimes \bbox{1}_{\rm E}
,U](P_1 \otimes \Sigma_{\rm E} )=\bbox{0}.
\label{the1pr1}
\end{equation}  
This relation and ${\cal T}_U(\rho+\rho^\prime)=\rho+\rho^\prime$
fulfill the requisite of Lemma 2 (with $\rho$ replaced by 
$\rho+\rho^\prime$), and we obtain 
\begin{equation}
[P_2\otimes \bbox{1}_{\rm E}
,U](P_2 \otimes \Sigma_{\rm E} )=\bbox{0}.
\label{the1pr2}
\end{equation}  
Using Eqs.~(\ref{the1pr1}) and (\ref{the1pr2}), we have 
$U=U(P_1 \otimes \Sigma_{\rm E})+U(P_2 \otimes \Sigma_{\rm E})
=\sum_{i=1,2}(P_i\otimes \bbox{1}_{\rm E})U(P_i \otimes \Sigma_{\rm E})$.
This implies that $U$ is written as $U=U_1\oplus U_2$
with $U_i\in {\cal U}({\cal H}_i) (i=1,2)$.

Now let us turn back to the classical example of preserving 
$p_1(x)$ and $p_2(x)$. 
We have seen that the transition 
$p(y|x)$ must occur within the sets $K_a$ and $K_b$ independently.
We can then consider each set separately. For example,
let us consider the conditional probability distributions
for $x\in K_a$, namely,
$p_s(x|x\in K_a)\equiv p_s(x)/\sum_{x\in K_a} p_s(x) (s=1,2)$.
The operation of $p(y|x)$ on the set $K_a$ should preserve
these two probability distributions. Then, 
if $p_1(x|x\in K_a)$ and $p_2(x|x\in K_a)$ are different, 
the above argument can be applied again, namely,
$K_a$ is separated into two subsets, within which 
the transition $p(y|x)$
should occur independently. These new sets and 
$K_b$ may 
be further separated into smaller ones by repeating
similar procedures. This refinement continues and 
should finally stop, as long as the set of all possible values 
$K_0$ is a finite set. In order to identify 
the final form of the refinement,
let us introduce the functions 
$f_s(x)\equiv p_s(x)/\sum_s p_s(x)$.
In a refinement process in which a subset 
$Y$ is divided into two subsets, 
the criteria of this division is 
whether $p_1(x|x\in Y)-p_2(x|x\in Y)$ is
positive or not. This function can be 
written in the form 
$(\alpha_1 f_1(x)-\alpha_2 f_2(x))\sum_s p_s(x)$.
Hence any two elements $x$ and $x^\prime$
that satisfy $f_s(x)=f_s(x^\prime)$ for all $s$ 
are always classified into the same subset.
If we write the final form as 
$K=\bigcup_l K^{(l)}$
with $K^{(l)}\cap K^{(l')}=\emptyset$ for $l\neq l'$,
$p_1(x|x\in K^{(l)})$ and $p_2(x|x\in K^{(l)})$ should be
identical for each subset $K^{(l)}$, since otherwise a 
further refinement would be possible. This condition
means that $f_s(x)=f_s(x^\prime)$ for all $s$ 
 and for any $x,x^\prime \in K^{(l)}$.
Therefore, the final form is the classification 
of the elements $x$ according to the 
set of values (a vector indexed by $s$) $\{f_s(x)\}$, and hence it is 
unique. This statement also holds for the 
cases when more than two probability distributions 
are preserved.

In quantum cases, we can similarly conduct the refinement 
of the decomposition of the Hilbert space into a direct sum
of subspaces by repeated uses of Theorem 1.
The final form of the decomposition, however, is not unique
in contrast to the classical cases.
One reason for this difference is that
the preservation of quantum states requires 
another type of conditions, which will be 
described in the next subsection. 

\subsection{Property of quantum signals}

In this subsection, we describe another basic theorem 
which applies when a state $\rho$ is preserved 
by an operation that affects two subspaces,
${\cal H}_1$ and ${\cal H}_2$, independently.
In order to preserve the off-diagonal part
$P_2\rho P_1$, the operation on ${\cal H}_1$
and that on ${\cal H}_2$ must satisfy a kind 
of `similarity'. This requirement is stated in the 
form of the following theorem.
\begin{theorem}
Let $P_1$ and $P_2$ be the projections onto orthogonal subspaces
${\cal H}_1$ and ${\cal H}_2$, respectively. 
Let $\rho$ be a density operator whose support is
${\cal H}_1\oplus {\cal H}_2$.
Suppose that $P_2\rho P_1\neq \bbox{0}$.
let ${\cal K}_1$ and ${\cal K}_2$ be 
the support and the image of $P_2\rho P_1$, respectively,
and ${\cal K}_i^\perp\equiv {\cal H}_i-{\cal K}_i (i=1,2)$.
Take the polar decomposition $P_2\rho P_1=WN$,
where $N$ is a positive operator on ${\cal K}_1$ and 
$W$ is a unitary operator from ${\cal K}_1$ to ${\cal K}_2$.
Then,
any pair of $U_i\in {\cal U}({\cal H}_i) (i=1,2)$ that satisfies 
${\cal T}_{U_1\oplus U_2}(\rho)=\rho$ 
can be written as $U_i= V_i \oplus \tilde{V}_i$, where 
$ V_i \in {\cal U}({\cal K}_i)$,
$\tilde{V}_i \in {\cal U}({\cal K}_i^\perp)$,
and 
\begin{equation}
V_2=(W\otimes {\bf 1}_{\rm E})
V_1(W^\dagger\otimes \Sigma_{\rm E}).
\label{th2}
\end{equation}
\end{theorem}

An intuitive explanation for this theorem is
as follows. The polar decomposition of 
$P_2\rho P_1$ means that 
it is written as $P_2\rho P_1=\sum_k a_k|k\rangle_{2}\;
\!{}_{1}\langle k |$, where $a_k$ are positive numbers, 
and $\{|k \rangle_i\}$ is a basis 
of ${\cal K}_i$ $(i=1,2)$. This implies that the coherence 
in $\rho$ is held in the pair 
$(|k \rangle_1,|k\rangle_2)$.
In order to retain this coherence, the operation 
${\cal T}_{U_1\oplus U_2}$ 
should not change this pairing relation,
 namely, if the operation
$V_1$ on ${\cal K}_1$
changes $|k \rangle_1$ to $|k^\prime \rangle_1$,
the operation $V_2$
on ${\cal K}_2$ should also change 
$|k\rangle_2$ to $|k^\prime\rangle_2$.
In addition, the change in the ancilla system ${\rm E}$
caused by the operation $V_1$ 
must be identical to that by $V_2$  in
order to avoid decoherence in the marginal state
for ${\cal H}_1\oplus {\cal H}_2$. 
Therefore, $V_1$ and $V_2$ must 
operate on ${\cal K}_1\otimes {\cal H}_{\rm E}$ and
 ${\cal K}_2\otimes {\cal H}_{\rm E}$ identically 
under the isomorphism $W$, which is implied by Eq.~(\ref{th2}).

Theorem 2 is proved as follows. 
Let us regard $N$ and $W$ as operators from 
${\cal H}_{12}\equiv {\cal H}_1\oplus {\cal H}_2$ to 
${\cal H}_{12}$
 by extending the domain and the range.
Note that $N:{\cal H}_{12}\rightarrow {\cal H}_{12}$ 
is a positive semidefinite operator with its support 
 ${\cal K}_1$ and its image  ${\cal K}_1$, and  
$W:{\cal H}_{12}\rightarrow {\cal H}_{12}$ is
 a partial isometry with its support ${\cal
K}_1$ and its image ${\cal
K}_2$.
The operator $W$ satisfies $W^2={\bf 0}$, 
$W^\dagger W$ is the projection onto ${\cal K}_1$,
and $WW^\dagger$ is the projection onto ${\cal K}_2$.
Let us define
\begin{equation}
P_{\pm}\equiv[W^\dagger W+WW^\dagger\pm(W+W^\dagger)]/2.
\label{the2pr1}
\end{equation}
These two operators are orthogonal projections since 
we can easily obtain $P_{\pm}^2=P_{\pm}$ and $P_+P_-=\bbox{0}$.
Note that 
$P_{+}+P_{-}$ is the projection onto ${\cal K}_1 \oplus {\cal
K}_2$.
Using these projections, define
\begin{equation}
O\equiv4(P_+\sqrt{N}P_+)^2-4(P_-\sqrt{N}P_-)^2.
\label{the2pr2}
\end{equation}
Substituting Eq.~(\ref{the2pr1}) and using  
relations such as 
$W^2=NW=\bbox{0}$ and $W^\dagger W\sqrt{N}=\sqrt{N}$,
we obtain $O= WN+NW^\dagger= P_2\rho P_1+P_1 \rho P_2$.
Since $O^2=WN^2W^\dagger+N^2$, the support of $O$ is 
${\cal K}_1 \oplus {\cal K}_2$.
Let us suppose that 
$U_1\in {\cal U}({\cal H}_1)$
and 
$U_2\in {\cal U}({\cal H}_2)$ satisfy
${\cal T}_{U_1\oplus U_2}(\rho)=\rho$.
Noting that 
\begin{equation}
[P_i\otimes \bbox{1}_{\rm E}
,U_1\oplus U_2](P_i \otimes \Sigma_{\rm E} )=\bbox{0} \;\;
(i=1,2),
\label{the2pr3}
\end{equation}  
we have 
\begin{equation}
{\cal T}_{U_1\oplus U_2}
(P_i\rho P_j)=P_i{\cal T}_{U_1\oplus U_2}(\rho)P_j
=P_i\rho P_j
\label{the2pr31}
\end{equation}  
for any $i=1,2$ and $j=1,2$.
From this relation, we have 
${\cal T}_{U_1\oplus U_2}(O)=O$.
The form of Eq.~(\ref{the2pr2}), together with the fact that
the support of $O$ coincides with the support of $P_++P_-$,
means that 
$P_+$ is the projection onto the space spanned 
by the eigenvectors of $O$ with
positive eigenvalues. Then, 
using Lemma 1, we obtain
\begin{equation}
[P_+\otimes \bbox{1}_{\rm E}
,U_1\oplus U_2](P_+ \otimes \Sigma_{\rm E} )=\bbox{0}.
\label{the2pr4}
\end{equation}  
Similarly, noting that ${\cal T}_{U_1\oplus U_2}(-O)=-O$
and that $P_-$ is the projection onto the space spanned 
by the eigenvectors of $-O$ with
positive eigenvalues, we have
\begin{equation}
[P_-\otimes \bbox{1}_{\rm E}
,U_1\oplus U_2](P_- \otimes \Sigma_{\rm E} )=\bbox{0}.
\label{the2pr5}
\end{equation}  
Combining Eqs.~(\ref{the2pr4})
 and (\ref{the2pr5}) with 
$W^\dagger W+WW^\dagger=P_++P_-$, we obtain
$[(W^\dagger W+WW^\dagger)\otimes \bbox{1}_{\rm E}
,U_1\oplus U_2]((W^\dagger W+WW^\dagger) \otimes \Sigma_{\rm E}
)=\bbox{0}$, or equivalently,
\begin{equation}
[W^\dagger W\otimes \bbox{1}_{\rm E}
,U_1](W^\dagger W \otimes \Sigma_{\rm E} )=\bbox{0}
\label{the2pr6}
\end{equation} 
and 
\begin{equation}
[WW^\dagger \otimes \bbox{1}_{\rm E}
,U_2](WW^\dagger  \otimes \Sigma_{\rm E} )=\bbox{0}.
\label{the2pr61}
\end{equation} 
From Eq.~(\ref{the2pr31}), we have 
${\cal T}_{U_1\oplus U_2}
(P_1\rho P_1)={\cal T}_{U_1}
(P_1\rho P_1)=P_1\rho P_1$.
Applying this and Eq.~(\ref{the2pr6}) to Lemma 2 
(note that the support of $P_1\rho P_1$ is 
${\cal H}_1={\cal K}_1\oplus {\cal K}_1^\perp$),
we obtain
\begin{equation}
[(P_1-W^\dagger W)\otimes \bbox{1}_{\rm E}
,U_1]((P_1-W^\dagger W) \otimes \Sigma_{\rm E} )=\bbox{0}.
\label{the2pr7}
\end{equation} 
Eqs.~(\ref{the2pr6}) and (\ref{the2pr7}) implies that 
$U_1$ can be written as 
$U_1= V_1 \oplus \tilde{V}_1$, where 
$ V_1 \in {\cal U}({\cal K}_1)$ and 
$\tilde{V}_1 \in {\cal U}({\cal K}_1^\perp)$
are related to $U_1$ as 
$V_1=U_1|{\cal K}_1$
and 
$\tilde{V}_1=U_1|{\cal K}_1^\perp$.
The same argument applies to
Eq.~(\ref{the2pr61}), leading to the conclusion that 
$U_2$ can be written as 
$U_2= V_2 \oplus \tilde{V}_2$, where 
$ V_2 \in {\cal U}({\cal K}_2)$ and 
$\tilde{V}_2 \in {\cal U}({\cal K}_2^\perp)$
are related to $U_2$ as 
$V_2=U_2|{\cal K}_2$
and 
$\tilde{V}_2=U_2|{\cal K}_2^\perp$.
Finally, combining Eqs.~(\ref{the2pr4})
 and (\ref{the2pr5}) with 
$W+W^\dagger=P_+-P_-$, we obtain
$[(W+W^\dagger)\otimes \bbox{1}_{\rm E}
,U_1\oplus U_2]((W+W^\dagger)\otimes \Sigma_{\rm E} )=\bbox{0}$.
Expanding this leads to
\begin{eqnarray}
(W\otimes \bbox{1}_{\rm E})
U_1(W^\dagger\otimes \Sigma_{\rm E} )
+(W^\dagger\otimes \bbox{1}_{\rm E})
U_2(W\otimes \Sigma_{\rm E} )
\nonumber \\
-U_1(W^\dagger W\otimes \Sigma_{\rm E} )
-U_2(WW^\dagger\otimes \Sigma_{\rm E} )
=\bbox{0}.
\label{the2pr8}
\end{eqnarray} 
Applying $P_2\otimes\bbox{1}_{\rm E}$ from the right
(and restricting the domain to ${\cal K}_2$),
we obtain Eq.~(\ref{th2}), which completes the proof.

\section{Structure of Hilbert space}
\label{sec:structure}

The requirement coming from Theorem 2 introduces a 
structure in the Hilbert space, which is more complicated
than a direct-sum decomposition into subspaces, namely,
some of the subspaces 
(e.g., ${\cal K}_1$ and ${\cal K}_2$)
are isometrically isomorphic 
through unitary operators 
(e.g., $W$)
connecting them. To handle 
such a structure in general, we introduce a way of decomposing
a Hilbert space ${\cal H}$ as follows. First, ${\cal H}$
is decomposed to a direct sum of its orthogonal subspaces
${\cal H}^{(1)},{\cal H}^{(2)},\ldots,{\cal H}^{(l_{\rm max})}$,
namely,
\begin{equation}
{\cal H}=\bigoplus_{l=1}^{l_{\rm max}}{\cal H}^{(l)}.
\label{defhl}
\end{equation} 
The size of each subspace is arbitrary. Then, each subspace
${\cal H}^{(l)}$
is further decomposed to a direct sum of its orthogonal subspaces
${\cal H}^{(l)}_1,{\cal H}^{(l)}_2,\ldots,
{\cal H}^{(l)}_{j^{(l)}_{\rm max}}$, namely,
\begin{equation}
{\cal H}^{(l)}=\bigoplus_{j=1}^{j^{(l)}_{\rm max}}{\cal H}_j^{(l)}.
\label{defhlj}
\end{equation} 
Here the subspaces $\{{\cal H}_j^{(l)}\}(j=1,2,\ldots,
j^{(l)}_{\rm max})$ are of the same size, and 
an isometrically isomorphic relation are defined among them
through a 
set of unitary operators $\{W^{(l)}_{j^\prime j}:
{\cal H}_j^{(l)}
\rightarrow {\cal H}_{j^\prime}^{(l)} \}$
satisfying 
$W^{(l)}_{k j}W^{(l)}_{j i}=
W^{(l)}_{k i}$. 
The numbers $l_{\rm max}$ and
$j^{(l)}_{\rm max}$ should satisfy
\begin{equation}
{\rm dim}\;{\cal H}=\sum_{l=1}^{l_{\rm max}}
j^{(l)}_{\rm max}{\rm dim}\;{\cal H}^{(l)}_1.
\end{equation}

The above decomposition can be completely specified 
by a set of partial isometries $\{W^{(l)}_{j^\prime j}\}$
acting on ${\cal H}$ satisfying the following three conditions,
\begin{equation}
W^{(l)\dagger}_{j^\prime j}=W^{(l)}_{jj^\prime},
\label{defw1}
\end{equation}
\begin{equation}
W^{(l^\prime)}_{j^\prime i^\prime}W^{(l)}_{i j}=
\delta_{l,l^\prime}\delta_{i,i^\prime}W^{(l)}_{j^\prime j},
\label{defw2}
\end{equation}
and 
\begin{equation}
\sum_{l=1}^{l_{\rm max}}\sum_{j=1}^{j^{(l)}_{\rm max}}W^{(l)}_{j j}
={\bf 1},
\label{defw3}
\end{equation}
where ${\bf 1}$ is the projection onto ${\cal H}$.
Given such $\{W^{(l)}_{j^\prime j}\}$, 
we can determine $\{{\cal H}_j^{(l)}\}$ as follows.
 From Eq.~(\ref{defw2}), we have 
$W^{(l^\prime)}_{j^\prime j^\prime}W^{(l)}_{j j}=
\delta_{l,l^\prime}\delta_{j,j^\prime}W^{(l)}_{j j}$. 
This means that $\{W^{(l)}_{j j}\}$ are projection operators
orthogonal to each other. If we take ${\cal H}^{(l)}_j$ as 
the support of $W^{(l)}_{j j}$, Eq.~(\ref{defw3}) assures that
Eqs.~(\ref{defhl}) and (\ref{defhlj}) are satisfied.
For $j\neq j^\prime$, the relation
$W^{(l)\dagger}_{j^\prime j}W^{(l)}_{j^\prime j}
=W^{(l)}_{j j}$ and 
$W^{(l)}_{j^\prime j}W^{(l)\dagger}_{j^\prime j}
=W^{(l)}_{j^\prime j^\prime}$ resulting from 
Eqs.~(\ref{defw1}) and (\ref{defw2}) means that
the support and the image of $W^{(l)}_{j^\prime j}$
are ${\cal H}^{(l)}_j$ and ${\cal H}^{(l)}_{j^\prime}$,
respectively. The map 
$W^{(l)}_{j^\prime j}:{\cal H}^{(l)}_j \rightarrow
{\cal H}^{(l)}_{j^\prime}$ is hence unitary and 
introduces an isometrically isomorphic relation
between ${\cal H}^{(l)}_j$ and ${\cal H}^{(l)}_{j^\prime}$.
The compatibility relation 
$W^{(l)}_{j^\prime i}W^{(l)}_{i j}
=W^{(l)}_{j^\prime j}$ coming from Eq.~(\ref{defw2}) 
assures that an isometrically isomorphic relation are 
defined among $\{{\cal H}_j^{(l)}\}(j=1,2,\ldots,
j^{(l)}_{\rm max})$.

The isomorphic relation among $\{{\cal H}_j^{(l)}\}$
 naturally defines an isomorphism (unitary map) $\Gamma^{(l)}$ 
from ${\cal H}^{(l)}$ to a tensor-product 
Hilbert space ${\cal H}^{(l)}_{\rm J}\otimes
{\cal H}^{(l)}_{\rm K}$, where 
${\rm dim}\;{\cal H}^{(l)}_{\rm J}=
j^{(l)}_{\rm max}$ and 
${\rm dim}\;{\cal H}^{(l)}_{\rm K}=
{\rm dim}\;{\cal H}_1^{(l)}
(={\rm dim}\;{\cal H}_2^{(l)}=\cdots)$. 
The unitary map
$\Gamma^{(l)}:{\cal H}^{(l)}\rightarrow
{\cal H}^{(l)}_{\rm J}\otimes
{\cal H}^{(l)}_{\rm K}$ is defined as follows. Take an 
arbitrary basis $\{|j\rangle^{(l)}_{\rm J}\}
(j=1,2,\ldots,j^{(l)}_{\rm max})$ for 
${\cal H}^{(l)}_{\rm J}$ and an arbitrary 
unitary operator $\Gamma_1^{(l)}$ from
${\cal H}_1^{(l)}$ to $|1\rangle^{(l)}_{\rm J}
\otimes{\cal H}^{(l)}_{\rm K}$. Then,
$\Gamma^{(l)}$ is given by
\begin{equation}
\Gamma^{(l)}=\sum_{j=1}^{j^{(l)}_{\rm max}}
(|j\rangle^{(l)}_{\rm J}\langle 1|\otimes {\bf 1}^{(l)}_{\rm K})
\Gamma_1^{(l)}W^{(l)}_{1j}.
\label{WtoGamma1}
\end{equation}
Form $\{\Gamma^{(l)}\}$, we can construct 
an isomorphism (unitary map) $\Gamma$ from ${\cal H}$ to
$\bigoplus_l{\cal H}^{(l)}_{\rm J}\otimes
{\cal H}^{(l)}_{\rm K}$  as 
\begin{equation}
\Gamma=\bigoplus_l \Gamma^{(l)}.
\label{WtoGamma2}
\end{equation}
Conversely, given a unitary map 
$\Gamma:{\cal H}\rightarrow
\bigoplus_l{\cal H}^{(l)}_{\rm J}\otimes
{\cal H}^{(l)}_{\rm K}$ we can construct 
a set of partial isometries $\{W^{(l)}_{j^\prime j}\}$
in ${\cal H}$ satisfying Eqs.~(\ref{defw1})--(\ref{defw3})
as follows. Take an 
arbitrary basis $\{|j\rangle^{(l)}_{\rm J}\}
(j=1,2,\ldots,j^{(l)}_{\rm max})$ for 
${\cal H}^{(l)}_{\rm J}$. Then, if we set
\begin{equation}
W^{(l)}_{j^\prime j}=\Gamma^\dagger
(|j^\prime\rangle^{(l)}_{\rm J}\langle j|
\otimes {\bf 1}^{(l)}_{\rm K})
\Gamma,
\label{GammatoW}
\end{equation}
Eqs.~(\ref{defw1})--(\ref{defw3}) are apparently satisfied.

In the above construction of $\{W^{(l)}_{j^\prime j}\}$ from 
$\Gamma$, we see that different decompositions,
$\{W^{(l)}_{j^\prime j}\}$ and  $\{\tilde{W}^{(l)}_{j^\prime j}\}$
for example, can
be derived  from the same $\Gamma$ due to the arbitrariness in the
choice  of basis $\{|j\rangle^{(l)}_{\rm J}\}$ . This implies that 
the two different sets $\{W^{(l)}_{j^\prime j}\}$ and 
$\{\tilde{W}^{(l)}_{j^\prime j}\}$ correspond to the same
structure in ${\cal H}$. The definition of $\Gamma$ also has 
 similar degeneracy, e.g., changing 
the order of the index $l$ merely alters 
the way of representation and does not change the 
structure itself. It is thus natural to define a 
{\it structure} in ${\cal H}$ as an equivalence class 
defined among the sets $\{\tilde{W}^{(l)}_{j^\prime j}\}$
or among the isometries $\Gamma$, in the following way.
Two decompositions specified by $\{W^{(l)}_{j^\prime j}\}$
and $\{\tilde{W}^{(l)}_{j^\prime j}\}$
are equivalent and correspond to the same structure if
\begin{equation}
\tilde{W}^{(P(l))}_{j^\prime j}=
\sum_{i,i^\prime}u^{(l)}_{j^\prime i^\prime}
W^{(l)}_{i^\prime i} u^{(l)*}_{j i},
\label{wunique}
\end{equation}
where $P(l)$ is a permutation of the index $l$
and $u^{(l)}_{i j}$ is the $(i,j)$ element of 
a unitary matrix $u^{(l)}$.
Two decompositions specified by 
$\Gamma:{\cal H}\rightarrow
\bigoplus_l{\cal H}^{(l)}_{\rm J}\otimes
{\cal H}^{(l)}_{\rm K}$ and 
$\tilde{\Gamma}:{\cal H}\rightarrow
\bigoplus_l\tilde{\cal H}^{(l)}_{\rm J}\otimes
\tilde{\cal H}^{(l)}_{\rm K}$ 
are equivalent and correspond to the same structure
if $\tilde{\Gamma}\Gamma^\dagger$
is written as 
\begin{equation}
\tilde{\Gamma}\Gamma^\dagger=
\bigoplus_lv^{(l)}_{\rm J}\otimes
v^{(l)}_{\rm K}
\label{Gammaunique}
\end{equation}
where $v^{(l)}_{\rm J}$ is a unitary map from
${\cal H}^{(l)}_{\rm J}$ to $\tilde{\cal H}^{(P(l))}_{\rm J}$,
and $v^{(l)}_{\rm K}$ is a unitary map from
${\cal H}^{(l)}_{\rm K}$ to $\tilde{\cal H}^{(P(l))}_{\rm K}$.

The relation among the definitions made so far is summarized as
follows. A structure $D$ is specified if a set 
$\{W^{(l)}_{j^\prime j}\}$ or a map $\Gamma$ is given.
Given a structure $D$, the set $\{W^{(l)}_{j^\prime j}\}$
and the map $\Gamma$ are not uniquely determined, 
and are only determined up to the conditions 
(\ref{wunique}) and (\ref{Gammaunique}).
The quantity $l_{\rm max}$ is uniquely determined,
and $\{j^{(l)}_{\rm max}\}$ are unique 
up to the permutation of the index $l$.

In the rest of the paper, we 
represent the isomorphic relation 
defined from $\Gamma:{\cal H}\rightarrow
\bigoplus_l{\cal H}^{(l)}_{\rm J}\otimes
{\cal H}^{(l)}_{\rm K}$
simply by
\begin{equation}
{\cal H}=\bigoplus_{l}{\cal H}^{(l)}_{\rm J}\otimes
{\cal H}^{(l)}_{\rm K}.
\end{equation}
An operator $A$ acting on ${\cal H}$ and 
an operator $A_{\rm JK}$ acting on 
$\bigoplus_l{\cal H}^{(l)}_{\rm J}\otimes
{\cal H}^{(l)}_{\rm K}$ is regarded as 
the same if 
\begin{equation}
A_{\rm JK}=\Gamma A \Gamma^\dagger
\end{equation}
holds. We also simply write this relation as
\begin{equation}
A_{\rm JK}=A,
\end{equation}
whenever the identity of $\Gamma$ is obvious in the 
context.

\section{Operation preserving a set of states}
\label{sec:main}

In this section, we give a solution to the problem formulated 
in Sec.~\ref{formulation}, namely, we identify the set 
${\cal U}_{\rm ND}$ given in Eq.~(\ref{defund}).
We first define a set of isometries ${\cal V}(D)$
associated with a structure $D$ in ${\cal H}_{\rm A}$, and define
an index $r(D)$ which gives the degree of refinement of $D$.
Then, 
 we apply Theorem 1 and
Theorem 2 repeatedly to refine the structure in ${\cal H}_{\rm A}$,
obtaining a series of structures $D_0,D_1,\ldots, D_{\rm fin}$
satisfying $r(D_0)<r(D_1)<\cdots<r(D_{\rm fin})$ and 
${\cal V}(D_0)\supset{\cal V}(D_1)
\supset{\cal V}(D_{\rm fin})\supset{\cal U}_{\rm ND}$.
It will be shown that under the final structure $D_{\rm fin}$,
the states $\{\rho_s\}$ have a simple form [Eq.~(\ref{decompositionrho})],
and we can easily identify the set ${\cal U}_{\rm ND}$. 

We first define a set of isometries ${\cal V}(D)$
associated with a structure $D$ in ${\cal H}_{\rm A}$.
 Let ${\cal
W}_D=\{W^{(l)}_{j^\prime j}\}$ be a set of isometries that specifies $D$.
With this notation, we define the set 
${\cal V}(D)$ as
\begin{eqnarray}
{\cal V}(D)\equiv\{U\in {\cal U}_{\rm all}|
{}^\forall W^{(l)}_{j^\prime j}\in {\cal W}_D,
\nonumber \\
U(W^{(l)}_{j^\prime j}\otimes \Sigma_{\rm E})
=(W^{(l)}_{j^\prime j}\otimes {\bf 1}_{\rm E})U
\}.
\label{defDbyW}
\end{eqnarray} 
This definition is consistent with the 
arbitrariness in the choice of 
${\cal W}_D$, namely, ${\cal V}(D)$
depends only on $D$.
Let $\Gamma_D:{\cal H}_{\rm A}\rightarrow 
\bigoplus_{l}{\cal H}^{(l)}_{\rm J}\otimes{\cal H}^{(l)}_{\rm K}$
be the isomorphism
determined from ${\cal W}_D$ through Eqs.~(\ref{WtoGamma1})
and (\ref{WtoGamma2}).
Under this isomorphism,
let $O_{\rm A}$ be an operator acting on 
${\cal H}_{\rm A}$ that is written as
\begin{equation}
O_{\rm A}=\bigoplus_l 
O_{\rm J}^{(l)}\otimes {\bf 1}^{(l)}_{\rm K},
\label{decompoa}
\end{equation}
where $O^{(l)}_{\rm J}$ operates on 
${\cal H}^{(l)}_{\rm J}$. 
Using Eq.~(\ref{GammatoW}), $O_{\rm A}$ is 
written as a linear combination
$O_{\rm A}=\sum_l\sum_{j,j^\prime}
c^{(l)}_{j^\prime j}W^{(l)}_{j^\prime j}$.
Hence, $U(O_{\rm A}\otimes \Sigma_{\rm E})
=(O_{\rm A}\otimes {\bf 1}_{\rm E})U$
holds for any $U \in {\cal V}(D)$.
Since $U$ satisfies this equation for any $O_{\rm A}$
in the form of Eq.~(\ref{decompoa}), we conclude
that any $U \in {\cal V}(D)$ can be written in 
 a simple form
\begin{equation}
U=\bigoplus_l {\bf 1}^{(l)}_{\rm J}\otimes U^{(l)}_{\rm KE},
\label{form_U_UKD}
\end{equation}
where $U^{(l)}_{\rm KE} \in {\cal U}({\cal H}^{(l)}_{\rm K})$.
Conversely, any isometry written in the form (\ref{form_U_UKD})
belongs to ${\cal V}(D)$, because any $W^{(l)}_{j^\prime j}$
has a form of $O_{\rm A}$ in Eq.~(\ref{decompoa}). 

Next, we introduce
an index $r(D)$ which represents the degree of refinement
of the structure $D$, defined as 
\begin{equation}
r(D)\equiv \frac{1}{2}\left(\sum_{l=1}^{l_{\rm max}}
j^{(l)}_{\rm max}\right)\left(\sum_{l=1}^{l_{\rm max}}
j^{(l)}_{\rm max}+1\right)-l_{\rm max}+1.
\label{defr}
\end{equation} 
This quantity takes an integer value in the following range
\begin{equation}
1\le r \le 
\frac{1}{2}({\rm dim}\;{\cal H}_{\rm A})
({\rm dim}\;{\cal H}_{\rm A}+1).
\label{rrange}
\end{equation}
This bound ensures that, 
when ${\rm dim}\;{\cal H}_{\rm A}$ is finite, any procedure of finding 
a series of structures with increasing degree of refinement
will halt within a finite number of steps.

The starting point of the refinement is 
to show that 
the trivial structure 
$D_0$ in ${\cal H}_{\rm A}$, for which $l_{\rm max}=1$, $j^{(1)}_{\rm
max}=1$, satisfies ${\cal U}_{\rm ND}\subset {\cal V}(D_0)$.
Applying $\rho_{\rm all}$ to Lemma 1 and 
noting Eq.~(\ref{Supprhoall}), we obtain 
\begin{equation}
[{\bf 1}_{\rm A}\otimes \bbox{1}_{\rm E}
,U]({\bf 1}_{\rm A} \otimes \Sigma_{\rm E} )=\bbox{0}
\end{equation}  
for any isometry $U\in {\cal U}_{\rm all}$
that satisfies ${\cal T}_U(\rho_{\rm all})=\rho_{\rm all}$.
Here ${\bf 1}_{\rm A}$ is the projection onto ${\cal H}_{\rm A}$.
This equation implies that the image of any $U\in {\cal U}_{\rm ND}$
 is a subspace of ${\cal H}_{\rm A}\otimes{\cal H}_{\rm E}$, 
namely, ${\cal U}_{\rm ND}\subset {\cal U}({\cal H}_{\rm A})$.
Since the set ${\cal W}_{D_0}$ consists of 
only one element, $W^{(1)}_{11}={\bf 1}_{\rm A}$, 
it is obvious from 
the definition (\ref{defDbyW}) that 
${\cal U}_{\rm ND}\subset {\cal V}(D_0)$.

Next, we state two lemmas to
 show that applying Theorem 1 and Theorem 2 
generally advances the refinement.

\begin{lemma}
Let $\Gamma:{\cal H}_{\rm A}\rightarrow
\bigoplus_l{\cal H}^{(l)}_{\rm J}\otimes
{\cal H}^{(l)}_{\rm K}$ be a unitary map 
that specifies a structure $D$.
Suppose that ${\cal U}_{\rm ND}\subset {\cal V}(D)$ and 
there exist $s\in S$, $l^\prime$, 
a pure state $|a\rangle \in {\cal H}^{(l^\prime)}_{\rm J}$,
and a unitary 
operator $V$ acting on ${\cal H}^{(l^\prime)}_{\rm J}$
such that for any $c\ge 0$,
\begin{eqnarray}
(|a\rangle\langle a|V\otimes {\bf 1}^{(l^\prime)}_{\rm K})
\rho_{s}
(V^\dagger|a\rangle\langle a|\otimes {\bf 1}^{(l^\prime)}_{\rm K})
\nonumber \\
\neq
c(|a\rangle\langle a|\otimes {\bf 1}^{(l^\prime)}_{\rm K})
\rho_{\rm all}
(|a\rangle\langle a|\otimes {\bf 1}^{(l^\prime)}_{\rm K}).
\end{eqnarray}
Then, there exists a structure $\tilde{D}$ such that
$r(\tilde{D})>r(D)$ and ${\cal U}_{\rm ND}\subset {\cal V}(\tilde{D})$.
\end{lemma}
For the proof, we actually construct $\tilde{D}$ assuming 
that $\rho\equiv
(|a\rangle\langle a|V\otimes {\bf 1}^{(1)}_{\rm K})
\rho_{s}
(V^\dagger|a\rangle\langle a|\otimes {\bf 1}^{(1)}_{\rm K})$
and $\rho^\prime\equiv
(|a\rangle\langle a|\otimes {\bf 1}^{(1)}_{\rm K})
\rho_{\rm all}
(|a\rangle\langle a|\otimes {\bf 1}^{(1)}_{\rm K})$
are different states
(here we have assumed that $l^\prime=1$, without loss of 
generality). 
Let ${\cal H}=|a\rangle\otimes
{\cal H}^{(1)}_{\rm K}$.
Eq.~(\ref{Supprhoall}) assures that ${\cal H}$ is the 
support of $\rho^\prime$, and hence is the support of 
$\rho+\rho^\prime$.
Then, using Theorem 1, we can find 
the decomposition ${\cal
H}={\cal H}_1\oplus {\cal H}_2$ where 
${\cal H}_1$ and ${\cal H}_2$ are nonzero subspaces.
Next, take a 
 basis $\{|j\rangle^{(l)}_{\rm J}\}
(j=1,2,\ldots,j^{(l)}_{\rm max})$ for 
${\cal H}^{(l)}_{\rm J}$ such that 
$|1\rangle^{(1)}_{\rm J}=|a\rangle$,
and construct a set ${\cal W}=\{W^{(l)}_{j^\prime j}\}$
by Eq.~(\ref{GammatoW}).
Let $P_1$ and $P_2$ be the projection onto 
${\cal H}_1$ and ${\cal H}_2$, respectively, and
define a new set $\tilde{\cal W}=\{\tilde{W}^{(l)}_{j^\prime j}\}$
as follows,
\begin{eqnarray}
\tilde{W}^{(1)}_{j^\prime j}&\equiv&
  W^{(1)}_{j^\prime 1}P_1W^{(1)}_{1 j}
\label{deftildew1}\\
\tilde{W}^{(l_{\rm max}+1)}_{j^\prime j}&\equiv&
  W^{(1)}_{j^\prime 1}P_2W^{(1)}_{1 j}
\\
\tilde{W}^{(l)}_{j^\prime j}&\equiv&
  W^{(l)}_{j^\prime j} \;\; (2\le l \le l_{\rm max}).
\label{deftildew3}
\end{eqnarray}
Noting that $P_1+P_2=W^{(1)}_{11}$, we can easily 
confirm that the conditions (\ref{defw1})--(\ref{defw3})
are satisfied by this new set $\tilde{\cal W}$, and 
hence $\tilde{\cal W}$ specifies a structure of 
${\cal H}_{\rm A}$. Let us denote this structure
by $\tilde{D}$.

The quantities 
$\tilde{l}_{\rm max}$ and $\tilde{j}^{(l)}_{\rm max}$ for 
$\tilde{\cal W}$ are related to 
$l_{\rm max}$ 
and $j^{(l)}_{\rm max}$ for ${\cal W}$ as
\begin{eqnarray}
\tilde{l}_{\rm max}&=&l_{\rm max}+1
\\
\tilde{j}^{(1)}_{\rm max}&=&\tilde{j}^{(l_{\rm max}+1)}_{\rm max}
=j^{(1)}_{\rm max}
\\
\tilde{j}^{(l)}_{\rm max}&=&
j^{(l)}_{\rm max}\;\; (2\le l \le l_{\rm max}).
\end{eqnarray}
Then, from Eq.~(\ref{defr}) and $j^{(1)}_{\rm max}\ge 1$, we have 
\begin{eqnarray}
\lefteqn{r(\tilde{D})-r(D)}
\nonumber \\
&=&
\frac{j^{(1)}_{\rm max}}{2}
\left(
2\sum_{l=1}^{l_{\rm max}}j^{(l)}_{\rm max}
+j^{(1)}_{\rm max}+1
\right)-1\ge 1.
\end{eqnarray}
Hence $r(\tilde{D})>r(D)$.

Since ${\cal U}_{\rm ND}\subset {\cal V}(D)$, 
any $U\in {\cal U}_{\rm
ND}$ can be written as 
$U=\bigoplus_l{\bf 1}^{(l)}_{\rm J}
\otimes U^{(l)}_{\rm KE}$
[Eq.~(\ref{form_U_UKD})]. From this form and the relations
${\cal T}_U(\rho_s)=\rho_s$ and 
${\cal T}_U(\rho_{\rm all})=\rho_{\rm all}$,
we have ${\cal T}_{U_0}(\rho)=\rho$ 
and ${\cal T}_{U_0}(\rho^\prime)=\rho^\prime$
where $U_0\equiv|a\rangle\langle a|\otimes 
U^{(1)}_{\rm KE}\in {\cal U}({\cal H})$.
Then, from Theorem 1, $U_0$ is written as 
$U_0=U_1\oplus U_2$
with $U_i\in {\cal U}({\cal H}_i) (i=1,2)$.
This form implies that
$U(P_i\otimes \Sigma_{\rm E})
=U_0(P_i\otimes \Sigma_{\rm E})
=(P_i\otimes {\bf 1}_{\rm E})U_0
=(P_i\otimes {\bf 1}_{\rm E})U$
for $i=1,2$.
Since $U\in {\cal V}(D)$,
$U(W^{(l)}_{j^\prime j}\otimes \Sigma_{\rm E})
=(W^{(l)}_{j^\prime j}\otimes {\bf 1}_{\rm E})U$
for any $W^{(l)}_{j^\prime j} \in {\cal W}$.
Combining these commuting relations and 
Eqs.~(\ref{deftildew1})--(\ref{deftildew3}),
we have 
\begin{equation}
U(\tilde{W}^{(l)}_{j^\prime j}\otimes \Sigma_{\rm E})
=(\tilde{W}^{(l)}_{j^\prime j}\otimes {\bf 1}_{\rm E})U
\end{equation}
for any $\tilde{W}^{(l)}_{j^\prime j} \in 
\tilde{\cal W}$. Hence 
$U\in {\cal V}(\tilde{D})$,
and we obtain
${\cal U}_{\rm ND}\subset {\cal V}(\tilde{D})$.
This completes the proof of Lemma 3.

\begin{lemma}
Let $\Gamma:{\cal H}_{\rm A}\rightarrow
\bigoplus_l{\cal H}^{(l)}_{\rm J}\otimes
{\cal H}^{(l)}_{\rm K}$ be a unitary map 
that specifies a structure $D$.
Suppose that ${\cal U}_{\rm ND}\subset {\cal V}(D)$ and 
there exist $s\in S$, $l^\prime$, $l^{\prime\prime}(\neq l')$,
a pure state
$|a\rangle$ in ${\cal H}^{(l^\prime)}_{\rm J}$,
and a pure state 
$|b\rangle$ in ${\cal H}^{(l^{\prime\prime})}_{\rm J}$
satisfying the following conditions,
\begin{equation}
{\rm Supp}((|a\rangle\langle a|
\otimes {\bf 1}^{(l^\prime)}_{\rm
K})\rho_{s}(|a\rangle\langle a|
\otimes {\bf 1}^{(l^{\prime})}_{\rm
K}))=|a\rangle\otimes{\cal H}^{(l^\prime)}_{\rm K},
\label{lemma4c1}
\end{equation}
\begin{equation}
{\rm Supp}((|b\rangle\langle b|
\otimes {\bf 1}^{(l^{\prime\prime})}_{\rm
K})\rho_{s}(|b\rangle\langle b|
\otimes {\bf 1}^{(l^{\prime\prime})}_{\rm
K}))=|b\rangle\otimes{\cal H}^{(l^{\prime\prime})}_{\rm K},
\label{lemma4c2}
\end{equation}
and 
\begin{equation}
(|b\rangle\langle b|
\otimes {\bf 1}^{(l^{\prime\prime})}_{\rm
K})\rho_{s}(|a\rangle\langle a|
\otimes {\bf 1}^{(l^{\prime})}_{\rm
K}) \neq {\bf 0}.
\label{lemma4c3}
\end{equation}
Then, there exists a structure $\tilde{D}$ such that
$r(\tilde{D})>r(D)$ and ${\cal U}_{\rm ND}\subset {\cal V}(\tilde{D})$.
\end{lemma}
For the proof, we actually construct $\tilde{D}$ assuming 
that conditions (\ref{lemma4c1})--(\ref{lemma4c3}) are 
satisfied for $l^\prime=1$ and
$l^{\prime\prime}=2$, without loss of 
generality.
Let ${\cal H}_1=|a\rangle\otimes
{\cal H}^{(1)}_{\rm K}$,
${\cal H}_2=|b\rangle\otimes
{\cal H}^{(2)}_{\rm K}$,
and $P_i$ be the projection onto 
${\cal H}_i(i=1,2)$.
Then, We can apply Theorem 2 by
choosing $\rho=(P_1+P_2)\rho_s(P_1+P_2)$,
and obtain
the decomposition
${\cal H}_i={\cal K}_i \oplus {\cal K}_i^\perp(i=1,2)$
 where 
${\cal K}_1$ and ${\cal K}_2$ are nonzero subspaces,
and the unitary operator $W:{\cal K}_1\rightarrow{\cal K}_2$.
Without loss of generality, we assume that 
${\rm dim}\;{\cal K}_1^\perp\ge{\rm dim}\;{\cal K}_2^\perp$.
Note that ${\cal K}_i^\perp(i=1,2)$ may be zero.
Next, take a 
 basis $\{|j\rangle^{(l)}_{\rm J}\}
(j=1,2,\ldots,j^{(l)}_{\rm max})$ for 
${\cal H}^{(l)}_{\rm J}$ ($l=1,2$) such that 
$|1\rangle^{(1)}_{\rm J}=|a\rangle$
and $|1\rangle^{(2)}_{\rm J}=|b\rangle$,
and construct a set ${\cal W}=\{W^{(l)}_{j^\prime j}\}$
by Eq.~(\ref{GammatoW}).
Let $Q_i$ and  $Q_i^\perp$ be the projection onto 
${\cal K}_i$ and ${\cal K}_i^\perp(i=1,2)$, respectively, and
define a new set $\tilde{\cal W}=\{\tilde{W}^{(l)}_{j^\prime j}\}$
as follows,
\begin{eqnarray}
\tilde{W}^{(1)}_{j^\prime j}&\equiv&
  W^{(1)}_{j^\prime 1}Q_1W^{(1)}_{1 j}
\label{deftildew4}\\
\tilde{W}^{(1)}_{j^\prime, \beta+j}&\equiv&
  W^{(1)}_{j^\prime 1}Q_1W^\dagger Q_2 W^{(2)}_{1 j}
\\
\tilde{W}^{(1)}_{\beta+j^\prime, j}&\equiv&
  W^{(2)}_{j^\prime 1}Q_2WQ_1W^{(1)}_{1 j}
\\
\tilde{W}^{(1)}_{\beta+j^\prime,\beta+j}&\equiv&
  W^{(2)}_{j^\prime 1}Q_2W^{(2)}_{1 j}
\\
\tilde{W}^{(l-1)}_{j^\prime j}&\equiv&
  W^{(l)}_{j^\prime j} \;\; (3\le l \le l_{\rm max})
\\
\tilde{W}^{(l_{\rm max})}_{j^\prime j}&\equiv&
  W^{(1)}_{j^\prime 1}Q_1^\perp W^{(1)}_{1 j}
\; ({\rm if} \; {\rm dim}\;{\cal K}_1^\perp\neq 0)
\\
\tilde{W}^{(l_{\rm max}+1)}_{j^\prime j}&\equiv&
  W^{(2)}_{j^\prime 1}Q_2^\perp W^{(2)}_{1 j}
\; ({\rm if} \; {\rm dim}\;{\cal K}_2^\perp\neq 0),
\label{deftildew10}
\end{eqnarray}
where $\beta\equiv j^{(1)}_{\rm max}$.
Noting that $Q_i+Q_i^\perp=W^{(i)}_{11}(i=1,2)$, we can easily 
confirm that the conditions (\ref{defw1})--(\ref{defw3})
are satisfied by this new set $\tilde{\cal W}$, and 
hence $\tilde{\cal W}$ specifies a structure of 
${\cal H}_{\rm A}$. Let us denote this structure
by $\tilde{D}$.

The quantities 
$\tilde{l}_{\rm max}$ and $\tilde{j}^{(l)}_{\rm max}$ for 
$\tilde{\cal W}$ are related to 
$l_{\rm max}$ 
and $j^{(l)}_{\rm max}$ for ${\cal W}$ as
\begin{eqnarray}
\tilde{l}_{\rm max}&=&l_{\rm max}-1+s_1+s_2
\\
\tilde{j}^{(1)}_{\rm max}&=&
j^{(1)}_{\rm max}
+j^{(2)}_{\rm max}
\\
\tilde{j}^{(l-1)}_{\rm max}&=&
j^{(l)}_{\rm max}\;\; (3\le l \le l_{\rm max})
\\
\tilde{j}^{(l_{\rm max})}_{\rm max}&=&
j^{(1)}_{\rm max}
\; ({\rm if} \; {\rm dim}\;{\cal K}_1^\perp\neq 0)
\\
\tilde{j}^{(l_{\rm max}+1)}_{\rm max}&=&
j^{(2)}_{\rm max}
\; ({\rm if} \; {\rm dim}\;{\cal K}_2^\perp\neq 0),
\end{eqnarray}
where $s_i=1$ if ${\rm dim}\;{\cal K}_i^\perp\neq 0$,
and $s_i=0$ if ${\rm dim}\;{\cal K}_i^\perp= 0$.
Then, from Eq.~(\ref{defr}), we have 
\begin{eqnarray}
\lefteqn{r(\tilde{D})-r(D)}
\nonumber \\
&=&
\frac{s}{2}
\left(
2\sum_{l=1}^{l_{\rm max}}j^{(l)}_{\rm max}
+s+1
\right)+1-s_1-s_2,
\end{eqnarray}
where $s\equiv s_1j^{(1)}_{\rm max}+s_2j^{(2)}_{\rm max}$.
Since $s\ge s_1+s_2\ge 0$, we obtain 
$r(\tilde{D})-r(D)\ge 1$.
Hence $r(\tilde{D})>r(D)$.

Since ${\cal U}_{\rm ND}\subset {\cal V}(D)$, 
any $U\in {\cal U}_{\rm
ND}$ can be written as 
$U=\bigoplus_l{\bf 1}^{(l)}_{\rm J}
\otimes U^{(l)}_{\rm KE}$. From this form and the relations
${\cal T}_U(\rho_s)=\rho_s$,
we have ${\cal T}_{U_1\oplus U_2}(\rho)=\rho$
where $U_1\equiv|a\rangle\langle a|\otimes 
U^{(1)}_{\rm KE}\in {\cal U}({\cal H}_1)$ and 
$U_2\equiv|b\rangle\langle b|\otimes 
U^{(2)}_{\rm KE}
\in {\cal U}({\cal H}_2)$.
Then, from Theorem 2, 
$U_i= V_i \oplus \tilde{V}_i$, where 
$ V_i \in {\cal U}({\cal K}_i)$,
$\tilde{V}_i \in {\cal U}({\cal K}_i^\perp) (i=1,2)$,
and 
$
V_2=(W\otimes {\bf 1}_{\rm E})
V_1(W^\dagger\otimes \Sigma_{\rm E})
$.
This form implies that, for $i=1,2$,
$U(Q_i\otimes \Sigma_{\rm E})
=V_i(Q_i\otimes \Sigma_{\rm E})
=(Q_i\otimes {\bf 1}_{\rm E})V_i
=(Q_i\otimes {\bf 1}_{\rm E})U$,
and 
$U(Q_i^\perp\otimes \Sigma_{\rm E})
=\tilde{V}_i(Q_i^\perp\otimes \Sigma_{\rm E})
=(Q_i^\perp\otimes {\bf 1}_{\rm E})\tilde{V}_i
=(Q_i^\perp\otimes {\bf 1}_{\rm E})U$.
We can also show that
$U(W\otimes \Sigma_{\rm E})
=V_2(W\otimes \Sigma_{\rm E})
=(W\otimes {\bf 1}_{\rm E})V_1
=(W\otimes {\bf 1}_{\rm E})U$
and 
$U(W^\dagger\otimes \Sigma_{\rm E})
=V_1(W^\dagger\otimes \Sigma_{\rm E})
=(W^\dagger\otimes {\bf 1}_{\rm E})V_2
=(W^\dagger\otimes {\bf 1}_{\rm E})U$.
Since $U\in {\cal V}(D)$,
$U(W^{(l)}_{j^\prime j}\otimes \Sigma_{\rm E})
=(W^{(l)}_{j^\prime j}\otimes {\bf 1}_{\rm E})U$
for any $W^{(l)}_{j^\prime j} \in {\cal W}$.
Combining these commuting relations and 
Eqs.~(\ref{deftildew4})--(\ref{deftildew10}),
we have 
\begin{equation}
U(\tilde{W}^{(l)}_{j^\prime j}\otimes \Sigma_{\rm E})
=(\tilde{W}^{(l)}_{j^\prime j}\otimes {\bf 1}_{\rm E})U
\end{equation}
for any $\tilde{W}^{(l)}_{j^\prime j} \in 
\tilde{\cal W}$. Hence 
$U\in {\cal V}(\tilde{D})$,
and we obtain
${\cal U}_{\rm ND}\subset {\cal V}(\tilde{D})$.
This completes the proof of Lemma 4.

Lemma 3 and Lemma 4 mean that starting from 
$D_0$, we can find a sequence $D_0,D_1,D_2,\ldots, D_n,\ldots $
that satisfies $r(D_0)<r(D_1)<r(D_2)<\ldots$ and 
${\cal U}_{\rm ND}\subset {\cal V}(D_n)$.
Since the integer value $r(D_n)$ has an upper bound
shown in Eq.~(\ref{rrange}), the sequence must 
end at some point. Let $D_{\rm fin}$ be the 
last one in the sequence,
and consider 
an isomorphism $\Gamma_{\rm fin}:{\cal H}_{\rm A}\rightarrow
\bigoplus_l{\cal H}^{(l)}_{\rm J}\otimes
{\cal H}^{(l)}_{\rm K}$ 
corresponding to $D_{\rm fin}$.
$D_{\rm fin}$ should not satisfy the 
prerequisites of Lemma 3 and Lemma 4.
From the prerequisite of Lemma 3,
wee see that $D_{\rm fin}$ satisfies the following:
 for any
$s\in S$, $l$, 
a pure state $|a\rangle \in {\cal H}^{(l)}_{\rm J}$,
and any unitary 
operator $V$ acting on ${\cal H}^{(l)}_{\rm J}$,
there exists $c\ge 0$ such that
\begin{eqnarray}
(|a\rangle\langle a|V\otimes {\bf 1}^{(l)}_{\rm K})
\rho_{s}
(V^\dagger|a\rangle\langle a|\otimes {\bf 1}^{(l)}_{\rm K})
\nonumber \\
=
c(|a\rangle\langle a|\otimes {\bf 1}^{(l)}_{\rm K})
\rho_{\rm all}
(|a\rangle\langle a|\otimes {\bf 1}^{(l)}_{\rm K}).
\label{nolemma3}
\end{eqnarray}
Let us fix $l$ and $|a\rangle$ for the moment.
 Because of Eq.~(\ref{Supprhoall}),
$Z\equiv{\rm Tr}[(|a\rangle\langle a|\otimes {\bf 1}^{(l)}_{\rm K})
\rho_{\rm all}
(|a\rangle\langle a|\otimes {\bf 1}^{(l)}_{\rm K})]\neq 0$.
Let us define a normalized 
density operator $\rho^{(l)}_{\rm K}$
acting on ${\cal H}^{(l)}_{\rm K}$ as 
\begin{equation}
\rho^{(l)}_{\rm K}\equiv
(\langle a|\otimes {\bf 1}^{(l)}_{\rm K})
\rho_{\rm all}
(|a\rangle\otimes {\bf 1}^{(l)}_{\rm K})
/Z.
\end{equation}
Eq.~(\ref{Supprhoall}) also assures that 
\begin{equation}
{\rm Supp}(\rho^{(l)}_{\rm K})={\cal H}^{(l)}_{\rm K}.
\label{SupprholK}
\end{equation}
The condition (\ref{nolemma3}) can be stated that, 
for any $s\in S$ and any unitary 
operator $V$, there exists $c^\prime\ge0$ such that
\begin{eqnarray}
(|a\rangle\langle a|V\otimes {\bf 1}^{(l)}_{\rm K})
\rho_{s}
(V^\dagger|a\rangle\langle a|\otimes {\bf 1}^{(l)}_{\rm K})
\nonumber \\
=
c^\prime|a\rangle\langle a|\otimes \rho^{(l)}_{\rm K}.
\label{nolemma31}
\end{eqnarray}
This is satisfied if and only if 
$\{\rho_s\}_{s\in S}$ are written in the form
\begin{equation}
({\bf 1}^{(l)}_{\rm J} \otimes {\bf 1}^{(l)}_{\rm K})
\rho_s
({\bf 1}^{(l)}_{\rm J} \otimes {\bf 1}^{(l)}_{\rm K})
=p^{(s,l)}\rho^{(s,l)}_{\rm J} \otimes \rho^{(l)}_{\rm K},
\label{resultnolemma3}
\end{equation}
where $p^{(s,l)}\ge 0$ and 
$\rho^{(s,l)}_{\rm J}$,
which is defined only when $p^{(s,l)}> 0$,
is a normalized 
density operator 
acting on ${\cal H}^{(l)}_{\rm J}$.
Note that $\rho^{(l)}_{\rm K}$
is independent 
of $s$.  

Next, let us consider the prerequisite 
(\ref{lemma4c1})--(\ref{lemma4c3}) of Lemma 4.
If Eq.~(\ref{lemma4c3}) is satisfied,
$(|a\rangle\langle a|
\otimes {\bf 1}^{(l^\prime)}_{\rm
K})\rho_{s}(|a\rangle\langle a|
\otimes {\bf 1}^{(l^{\prime})}_{\rm
K})\neq {\bf 0}$. Then, 
the form (\ref{resultnolemma3}) and Eq.~(\ref{SupprholK})
implies that  Eq.~(\ref{lemma4c1}) is also satisfied.
Similarly, Eq.~(\ref{lemma4c2}) is also satisfied and
all the prerequisite is met. Therefore,
the condition that $D_{\rm fin}$ should not satisfy the 
prerequisite of Lemma 4 means that
\begin{equation}
({\bf 1}^{(l^\prime)}_{\rm J} \otimes {\bf 1}^{(l^\prime)}_{\rm K})
\rho_s
({\bf 1}^{(l)}_{\rm J} \otimes {\bf 1}^{(l)}_{\rm K})
={\bf 0}
\label{resultnolemma4}
\end{equation}
for any $l$ and $l^\prime(\neq l)$.

Now we can state the main conclusion of this paper.
From Eqs.~(\ref{Supprhos}), (\ref{resultnolemma3}),
and (\ref{resultnolemma4}), we conclude that $\rho_s$
is written as 
\begin{equation}
\rho_s
=\bigoplus_l p^{(s,l)}\rho^{(s,l)}_{\rm J} \otimes \rho^{(l)}_{\rm K},
\label{decompositionrho}
\end{equation}
under the decomposition 
of their support ${\cal H}_{\rm A}$,
\begin{equation}
{\cal H}_{\rm A}=\bigoplus_{l}{\cal H}^{(l)}_{\rm J}\otimes
{\cal H}^{(l)}_{\rm K},
\label{decompositionha}
\end{equation}
which corresponds to $D_{\rm fin}$.
Here $\rho^{(s,l)}_{\rm J}$ and $\rho^{(l)}_{\rm K}$
 are normalized
density  operators acting on ${\cal H}^{(l)}_{\rm J}$ and 
${\cal H}^{(l)}_{\rm K}$,
respectively,  and
$p^{(s,l)}$ is the probability for the state $\rho_s$
to be  in the
subspace
${\cal H}^{(l)}_{\rm J}\otimes
{\cal H}^{(l)}_{\rm K}$. Note that $\rho^{(l)}_{\rm K}$
is independent 
of $s$.  
Since ${\cal U}_{\rm ND}\subset {\cal V}(D_{\rm fin})$,
any $U \in {\cal U}_{\rm ND}$ should be written as
\begin{equation}
U=\bigoplus_{l}{\bf 1}^{(l)}_{\rm J}\otimes
U^{(l)}_{\rm KE},
\label{main1}
\end{equation}
where $U^{(l)}_{\rm KE} \in {\cal U}({\cal H}^{(l)}_{\rm K})$.
It is obvious that 
$U^{(l)}_{\rm KE}$ must obey 
\begin{equation}
 \mbox{Tr}_{\rm E}[U^{(l)}_{\rm KE}
(\rho^{(l)}_{\rm K}\otimes \Sigma_{\rm E} )
U^{(l)\dagger}_{\rm KE}]=\rho^{(l)}_{\rm K}.
\label{main2}
\end{equation}
The condition expressed by (\ref{main1}) and (\ref{main2}) 
together is an equivalent condition for the condition 
${\cal T}_U(\rho_s)=\rho_s$, since the sufficiency is apparently satisfied.

The condition (\ref{main1}), which is applied for an 
isometry $U:{\cal H}_{\rm A}\otimes|u\rangle_{\rm E}
\rightarrow {\cal H}_{\rm A}^\prime\otimes{\cal H}_{\rm E}$,
can be rewritten in the form that applies to a unitary operator
acting on ${\cal H}_{\rm A}^\prime\otimes {\cal H}_{\rm E}$
as follows.
Any unitary operator $U$ acting on ${\cal H}_{\rm A}^\prime\otimes {\cal
H}_{\rm E}$ that preserves $\{\rho_s\}_{s\in S}$ is expressed 
in the following
form,
\begin{equation}
U(\bbox{1}_{\rm A}\otimes \Sigma_{\rm E})
=\bigoplus_l\bbox{1}^{(l)}_{\rm J}\otimes U^{(l)}_{\rm
KE}(\bbox{1}^{(l)}_{\rm K}\otimes \Sigma_{\rm E}), 
\label{main3}
\end{equation}
where $U^{(l)}_{\rm KE}$ are unitary operators acting on the combined
space ${\cal H}^{(l)}_{\rm K}\otimes {\cal H}_{\rm E}$.

From the decomposition (\ref{decompositionrho}), we can classify 
the degrees of freedom of the system into three types ---
(a) The index $l$. The information on $s$ is stored 
classically, since there are no off-diagonal elements
and everything is expressed by the probability distribution $p^{(s,l)}$.
The operation $U$, which preserves $\{\rho_s\}$,
 must act independently on each subspace
${\cal H}^{(l)}_{\rm J}\otimes
{\cal H}^{(l)}_{\rm K}$. With such $U$ one can establish classical 
correlations between the system and the ancilla through $l$, but 
not quantum correlations. (b) The inner degree of freedom for each 
${\cal H}_{\rm J}^{(l)}$. The information on $s$ is stored
nonclassically, 
in the sense that there are nonvanishing off-diagonal elements
for any nontrivial observables. The operation $U$ must not act on 
this degree of freedom. (c) The inner degree of freedom for each 
${\cal H}_{\rm K}^{(l)}$. No information on $s$ is stored here. The
operation
$U$ can do anything as long as it leaves the system in the known 
state $\rho^{(l)}_{\rm K}$. For example, one can establish quantum
correlation between the system and the ancilla. In short, the principle
derived here  is stated as follows. In order to preserve the state of a
system,  no access is allowed to the part with quantum information,
classical  access is allowed to the part with classical information, and
quantum access is allowed to the part with no information.

\section{Properties of structure}
\label{sec:properties}

In the last section, we introduced a procedure to 
actually construct a structure $D_{\rm fin}$, and stated
the principle for the operations preserving $\{\rho_s\}$
using $D_{\rm fin}$. In this section, we will show that 
the structure $D_{\rm fin}$ derived from the procedure 
is unique. We will also give a criteria of determining
 whether a given structure is equivalent to $D_{\rm fin}$
or not, without doing the procedure in Sec.~\ref{sec:main}.
 
We first define a property of structure called `maximal',
which is, as will soon be shown, the property possessed by 
 $D_{\rm fin}$.

\begin{definition}
Let  $\Gamma:{\cal H}_{\rm A}\rightarrow
\bigoplus_l{\cal H}^{(l)}_{\rm J}\otimes
{\cal H}^{(l)}_{\rm K}$ be a unitary isomorphism
corresponding to a structure $D$.
We call $D$ maximal if the following three conditions are met.

{\rm (i)}
${\Gamma}\rho_s{\Gamma}^\dagger$ is written as
\begin{equation}
{\Gamma}\rho_s{\Gamma}^\dagger
=\bigoplus_l {p}^{(s,l)}\rho^{(s,l)}_{\rm J} \otimes
\rho^{(l)}_{\rm K},
\label{rhodecompmax}
\end{equation}
where $\rho^{(s,l)}_{\rm J}$ and $\rho^{(l)}_{\rm K}$
 are normalized
density  operators acting on ${\cal H}^{(l)}_{\rm J}$ and 
${\cal H}^{(l)}_{\rm K}$,
respectively.

{\rm (ii)}
If a projection $P:{\cal H}^{(l)}_{\rm
J}\rightarrow {\cal H}^{(l)}_{\rm J}$ satisfies 
\begin{equation}
P {p}^{(s,l)}\rho^{(s,l)}_{\rm J}
={p}^{(s,l)}\rho^{(s,l)}_{\rm J}
P
\end{equation}
for all $s\in S$, then $P={\bf 1}^{(l)}_{\rm
J}$ or $P={\bf 0}$.

{\rm (iii)}
No unitary operator $V:{\cal H}^{(l)}_{\rm
J}\rightarrow {\cal H}^{(l^\prime)}_{\rm J}$ $(l\neq
l^\prime)$ exists that satisfies 
\begin{equation}
V  {p}^{(s,l)}\rho^{(s,l)}_{\rm J}
=\alpha {p}^{(s,l^\prime)}\rho^{(s,l^\prime)}_{\rm J}
V
\end{equation}
for all $s\in S$ and for a positive number $\alpha$.
\end{definition}

We will then prove that  the structure derived 
and used in the previous section satisfies the above 
conditions, namely, 
\begin{lemma}
Any structure $D_{\rm fin}$ derived by the procedure 
in Sec.~\ref{sec:main} is maximal.
\label{finismax}
\end{lemma}
Let  $\Gamma_{\rm fin}:{\cal H}_{\rm A}\rightarrow
\bigoplus_l{\cal H}^{(l)}_{\rm J}\otimes
{\cal H}^{(l)}_{\rm K}$ be an isomorphism
corresponding to $D_{\rm fin}$.
The condition (i) in Definition 1 is apparently satisfied. 
For the condition (ii), suppose that a projection 
$P:{\cal H}^{(l)}_{\rm
J}\rightarrow {\cal H}^{(l)}_{\rm J}$ satisfies
$P{p}^{(s,l)}\rho^{(s,l)}_{\rm J}
={p}^{(s,l)}\rho^{(s,l)}_{\rm J}P$ for any $s\in S$. 
Construct an isometry $U_1\in {\cal U}({\cal H}_{\rm A})$ 
 such that it operates on ${\cal H}^{(l)}$ as  
\begin{equation}
U_1[({\bf 1}^{(l)}_{\rm J}\otimes {\bf 1}^{(l)}_{\rm K})\otimes
|u\rangle_{\rm E}\langle u|]=(P\otimes {\bf 1}^{(l)}_{\rm K})\otimes
|u^\perp\rangle_{\rm E}\langle u| +[({\bf 1}_{\rm J}^{(l)}-P)\otimes {\bf
1}^{(l)}_{\rm K}]
\otimes |u\rangle_{\rm E}\langle u|,
\end{equation}
where $|u^\perp\rangle_{\rm E}$ is a state orthogonal to 
$|u\rangle_{\rm E}$, and $U_1$ leaves the other subspaces 
unaltered.
It is easy to show that $U_1\in {\cal U}_{\rm ND}$ using the 
relation $P{p}^{(s,l)}\rho^{(s,l)}_{\rm J}
={p}^{(s,l)}\rho^{(s,l)}_{\rm J}P$. This 
means that $U_1$ should be written in the form of Eq.~(\ref{main1}),
which is only possible when $P={\bf 1}^{(l)}_{\rm
J}$ or $P= {\bf 0}$. 
For the condition (iii),
we will show that the existence of $V$ leads to a contradiction.
Without loss of generality, assume that there exists 
a unitary operator $V:{\cal H}^{(1)}_{\rm
J}\rightarrow {\cal H}^{(2)}_{\rm J}$
that satisfies 
${p}^{(s,1)}V\rho^{(s,1)}_{\rm J}V^\dagger
=\alpha {p}^{(s,2)}\rho^{(s,2)}_{\rm J}$.
 We can construct an isometry 
$U_2\in {\cal U}_{\rm ND}$ in the following way.
Let ${\cal H}^{(1)}_{\rm E}$ and ${\cal H}^{(2)}_{\rm E}$
be orthogonal subspaces of ${\cal H}_{\rm E}$ that are also 
orthogonal to $|u\rangle_{\rm E}$.
There exists an isometry $V^{(21)}_{\rm KE} 
:{\cal H}^{(1)}_{\rm K}\otimes |u\rangle_{\rm E}
\rightarrow {\cal H}^{(2)}_{\rm K}\otimes {\cal H}^{(2)}_{\rm E}$
satisfying 
$ \mbox{Tr}_{\rm E}[V^{(21)}_{\rm KE}
(\rho^{(1)}_{\rm K}\otimes \Sigma_{\rm E} )
V^{(21)\dagger}_{\rm KE}]=\rho^{(2)}_{\rm K}$.
Physically, a simple example is 
the operation that discards the input state
away and prepares the system in $\rho^{(2)}_{\rm K}$.
Similarly, let $V^{(12)}_{\rm KE} 
:{\cal H}^{(2)}_{\rm K}\otimes |u\rangle_{\rm E}
\rightarrow {\cal H}^{(1)}_{\rm K}\otimes {\cal H}^{(1)}_{\rm E}$
 be an isometry satisfying 
 $\mbox{Tr}_{\rm E}[V^{(12)}_{\rm KE}
(\rho^{(2)}_{\rm K}\otimes \Sigma_{\rm E} )
V^{(12)\dagger}_{\rm KE}]=\rho^{(1)}_{\rm K}$.
Then, we can construct a unitary operator $U_2$ 
such that it acts on ${\cal H}^{(1)}\oplus{\cal H}^{(2)}$
as
\begin{eqnarray}
U_2[(P^{(1)}+ P^{(2)})\otimes \Sigma_{\rm E}]
&=&\beta (P^{(1)}\otimes \Sigma_{\rm
E}+V^\dagger\otimes V^{(12)}_{\rm
KE})
\nonumber \\
&&+\sqrt{1-\beta^2}(V\otimes V^{(21)}_{\rm KE}+
P^{(2)}\otimes \Sigma_{\rm E}),
\end{eqnarray}
where $P^{(l)}\equiv {\bf 1}^{(l)}_{\rm J}\otimes
{\bf 1}^{(l)}_{\rm K}$ $(l=1,2)$ and 
$\beta=\sqrt{\alpha/
(1+\alpha)}$,
and it does nothing on the other subspaces with $l>2$.
It is then easy to show that $U_2\in {\cal U}_{\rm ND}$ using the 
relation ${p}^{(s,1)}V\rho^{(s,1)}_{\rm J}V^\dagger
=\alpha {p}^{(s,2)}\rho^{(s,2)}_{\rm J}$.
On the other hand, 
because of the cross terms 
$V^\dagger\otimes V^{(12)}_{\rm
KE}$ and
$V\otimes V^{(21)}_{\rm KE}$,
  $U_2$ cannot be 
written in the form of Eq.~(\ref{main1}), leading to 
a contradiction. The lemma is thus proved.

It is convenient to give a lemma showing that the conditions 
(ii) and (iii) in Definition 1 are equivalent to 
slightly stronger conditions.

\begin{lemma}
Let $\Gamma:{\cal H}_{\rm A}\rightarrow
\bigoplus_l{\cal H}^{(l)}_{\rm J}\otimes
{\cal H}^{(l)}_{\rm K}$ be a unitary isomorphism  
that corresponds to a maximal structure $D$, and 
${\Gamma}\rho_s{\Gamma}^\dagger
=\bigoplus_l {p}^{(s,l)}\rho^{(s,l)}_{\rm J} \otimes
\rho^{(l)}_{\rm K}$.
Then, 

{\rm (ii$'$)}
If an operator $\Lambda: {\cal H}^{(l)}_{\rm
J}\rightarrow {\cal H}^{(l)}_{\rm J}$ satisfies 
\begin{equation}
\Lambda {p}^{(s,l)}\rho^{(s,l)}_{\rm J}
=\beta {p}^{(s,l)}\rho^{(s,l)}_{\rm J}
\Lambda
\label{prooflemma6-1}
\end{equation}
for all $s\in S$ and for a complex number $\beta$, 
then $\Lambda=c{\bf
1}^{(l)}_{\rm J}$, where $c$ is a complex number.
Especially, $\Lambda={\bf 0}$ when $\beta\neq 1$.

{\rm (iii$'$)}
If an operator $\Lambda:{\cal H}^{(l)}_{\rm
J}\rightarrow {\cal H}^{(l^\prime)}_{\rm J}$ $(l\neq
l^\prime)$ satisfies 
\begin{equation}
\Lambda  {p}^{(s,l)}\rho^{(s,l)}_{\rm J}
=\alpha {p}^{(s,l^\prime)}\rho^{(s,l^\prime)}_{\rm J}
\Lambda
\label{prooflemma6-2}
\end{equation}
for all $s\in S$ and for a positive number $\alpha$,
then $\Lambda={\bf 0}$.
\label{Schur}
\end{lemma}
First, we prove (ii$'$). 
If $\Lambda$ is invertible in
${\cal H}^{(l)}_{\rm
J}$, operating $\Lambda^{-1}$ from the left 
and taking the trace for both sides of 
Eq.~(\ref{prooflemma6-1}), we have ${p}^{(s,l)}=\beta {p}^{(s,l)}$
for any $s$ and hence $\beta=1$. $\Lambda$ is thus not invertible
if $\beta\neq 1$.
Let $c$ be an eigenvalue of $\Lambda$
when $\beta=1$, and let $c=0$ when $\beta\neq 1$.
Define $\Lambda^\prime\equiv \Lambda-c{\bf 1}^{(l)}_{\rm
J}$. Then, 
 $\Lambda^\prime$ is not invertible in
${\cal H}^{(l)}_{\rm
J}$ for any value of $\beta$.
From Eq.~(\ref{prooflemma6-1}), 
we have $\Lambda^\prime {p}^{(s,l)}\rho^{(s,l)}_{\rm J}
=\beta {p}^{(s,l)}\rho^{(s,l)}_{\rm J}
\Lambda^\prime$.
Let $P$ be the projection onto 
${\rm Ker} \Lambda^\prime$, the kernel of $\Lambda^\prime$. For any
vector 
$|a\rangle \in {\rm Ker} \Lambda^\prime$, 
$\Lambda^\prime{p}^{(s,l)}\rho^{(s,l)}_{\rm J}|a\rangle
=\beta{p}^{(s,l)}\rho^{(s,l)}_{\rm J}\Lambda^\prime|a\rangle=0$ and hence 
${p}^{(s,l)}\rho^{(s,l)}_{\rm J}|a\rangle \in {\rm Ker} \Lambda^\prime$.
We thus have $P{p}^{(s,l)}\rho^{(s,l)}_{\rm J}
={p}^{(s,l)}\rho^{(s,l)}_{\rm J}P$.
Since $\Lambda^\prime$ 
is not invertible
in ${\cal H}^{(l)}_{\rm J}$, $P\neq {\bf 0}$.
From (ii) in Definition 1, we have $P={\bf 1}^{(l)}_{\rm
J}$ and $\Lambda^\prime={\bf 0}$, 
hence $\Lambda=c{\bf 1}^{(l)}_{\rm J}$.
For the proof of (iii$'$), suppose that 
Eq.~(\ref{prooflemma6-2}) holds. Together with its 
Hermite conjugate, we have 
$\Lambda^\dagger \Lambda  {p}^{(s,l)}\rho^{(s,l)}_{\rm J}
= {p}^{(s,l)}\rho^{(s,l)}_{\rm J}\Lambda^\dagger \Lambda$
for all $s$. From (ii$'$), 
$\Lambda^\dagger \Lambda=c{\bf 1}^{(l)}_{\rm J}$.
A similar argument gives 
$\Lambda \Lambda^\dagger=c^\prime{\bf 1}^{(l^\prime)}_{\rm J}$.
If $\Lambda\neq {\bf 0}$, $c> 0$ and $\Lambda/\sqrt{c}:
{\cal H}^{(l)}_{\rm
J}\rightarrow {\cal H}^{(l^\prime)}_{\rm J}$ is unitary,
but this conflicts with (iii) in Definition 1.
Hence $\Lambda= {\bf 0}$.

The conditions (ii) and (iii) for maximal structures 
have a simple meaning when we consider the algebra 
over $C$ generated by the set 
of operators $\{\rho_s\}_{s\in S}$. Let us denote this 
algebra by ${\cal X}$. ${\cal H}_{\rm A}$ is then regarded
as a ${\cal X}$-module.
Let  $\Gamma:{\cal H}_{\rm A}\rightarrow
\bigoplus_l{\cal H}^{(l)}_{\rm J}\otimes
{\cal H}^{(l)}_{\rm K}$ be an isomorphism
corresponding to a maximal structure $D$.
Let us write a
diagonalization of $\rho^{(l)}_{\rm K}$ as
\begin{equation}
\rho^{(l)}_{\rm K}=\sum_k q^{(l)}_k|a_k\rangle^{(l)}_{\rm K}
\langle a_k|,
\end{equation}
where $\{|a_k\rangle^{(l)}_{\rm K}\}$ 
$(k=1,2,\ldots,{\rm dim}{\cal H}^{(l)}_{\rm K})$ 
is a basis of ${\cal H}^{(l)}_{\rm K}$, and 
$q^{(l)}_k>0$ since ${\rm Supp}(\rho_{\rm all})={\cal H}_{\rm A}$.
 Consider a direct-sum decomposition 
${\cal H}_{\rm A}=\bigoplus_l(\bigoplus_k {\cal H}^{(l,k)})$,
where ${\cal H}^{(l,k)}={\cal H}^{(l)}_{\rm J}\otimes
|a_k\rangle^{(l)}_{\rm K}$ under the isomorphism $\Gamma$. ${\cal
H}^{(l,k)}$ are then 
 ${\cal X}$-submodules, namely, $A|x\rangle\in {\cal H}^{(l,k)}$
for any $A\in {\cal X}$ and for any $|x\rangle\in {\cal H}^{(l,k)}$.
The condition (ii) or (ii$'$) implies that
${\cal H}^{(l,k)}$ is simple, namely,
it has no submodules other than zero and ${\cal H}^{(l,k)}$ itself.
The condition (iii) or (iii$'$) means that two submodules
${\cal H}^{(l,k)}$ and ${\cal H}^{(l^\prime,k^\prime)}$
with $l\neq l^\prime$ are not ${\cal X}$-isomorphic.
To show it, suppose that 
${\cal H}^{(l,k)}$ and ${\cal H}^{(l^\prime,k^\prime)}$
 are ${\cal X}$-isomorphic, namely, there exists a 
linear invertible map $\Lambda:{\cal H}^{(l,k)}
\rightarrow {\cal H}^{(l^\prime,k^\prime)}$ satisfying 
$\Lambda A|x\rangle= A \Lambda |x\rangle$ for 
any $A\in {\cal X}$ and for any $|x\rangle\in {\cal H}^{(l,k)}$.
$\Gamma \Lambda \Gamma^\dagger$ is 
 written as 
$\Gamma \Lambda \Gamma^\dagger=\Lambda^\prime\otimes 
|a_{k^\prime}\rangle^{(l^\prime)}_{\rm K}
{}^{(l)}_{\rm K}\langle a_k |$, where 
$\Lambda^\prime$ is a nonzero operator
from ${\cal H}^{(l)}_{\rm J}$ to 
${\cal H}^{(l^\prime)}_{\rm J}$.
Since $\Lambda \rho_s= \rho_s \Lambda$
for any $s \in S$,
we have 
\begin{equation}
q_k^{(l)} \Lambda^\prime p^{(s,l)}  \rho^{(s,l)}_{\rm J}
=q_{k^\prime}^{(l^\prime)} p^{(s,l^\prime)} \rho^{(s,l^\prime)}_{\rm J}
\Lambda^\prime.
\label{isomorphic1}
\end{equation}
Lemma \ref{Schur} implies that this happens only when 
$l=l^\prime$.

While ${\cal H}^{(l,k)}$ and ${\cal H}^{(l^\prime,k^\prime)}$
 are not ${\cal X}$-isomorphic when $l\neq l^\prime$, 
${\cal H}^{(l,k)}$ and ${\cal H}^{(l,k^\prime)}$
are ${\cal X}$-isomorphic only when 
$q_k^{(l)}=q_{k^\prime}^{(l)}$, and not 
${\cal X}$-isomorphic when $q_k^{(l)}\neq q_{k^\prime}^{(l)}$.
It may be convenient if we can construct an algebra 
$\tilde{\cal X}$ such that 
${\cal H}^{(l,k)}$ and ${\cal H}^{(l^\prime,k^\prime)}$
 are $\tilde{\cal X}$-isomorphic iff $l=l^\prime$. 
We will show that such an algebra can be constructed by 
`normalizing' $\{\rho_s\}$ relative to $\rho_{\rm all}$.
First, take a decomposition of ${\cal H}_{\rm A}$ into 
simple ${\cal X}$-submodules, 
${\cal H}_{\rm A}=\bigoplus_m(\bigoplus_i{\cal H}_{mi})$,
where ${\cal H}_{mi}$ and ${\cal H}_{m^\prime i^\prime}$
are ${\cal X}$-isomorphic iff $m=m^\prime$.
Let $P_{mi}$ be the projection onto ${\cal H}_{mi}$,
and $P_{m}\equiv \sum_i P_{mi}$.
Then, define $\tilde\rho_s$ as 
\begin{equation}
\tilde\rho_s=\sum_{m,i} \rho_s  P_{mi}[{\rm Tr}(\rho_{\rm
all}P_{mi})]^{-1} =\sum_m
[{\rm Tr}(\rho_{\rm all}P_{m1})]^{-1}  \rho_s  
P_{m}
\label{def_tilde_rho}
\end{equation}
where we have used the fact that 
${\rm Tr}(\rho_{\rm all}P_{mi})$ is independent of $i$ 
since $\rho_{\rm all}\in {\cal X}$.
Let $\tilde{\cal X}$ be the algebra 
over $C$ generated by the set 
of operators $\{\tilde\rho_s\}_{s\in S}$.
This definition is independent of the choice of the 
decomposition ${\cal H}_{\rm A}=\bigoplus_m(\bigoplus_i{\cal H}_{mi})$. 
To prove it, take another decomposition 
${\cal H}_{\rm A}=\bigoplus_m(\bigoplus_i{\cal H}^\prime_{mi})$
and define $P^\prime_{mi}$ and $P^\prime_m$ in the same way as before.
The number of submodules are the same in the two decompositions, 
and we can make ${\cal H}_{mi}$ and ${\cal H}^\prime_{mi}$ 
be ${\cal X}$-isomorphic by appropriately arranging the order of
summation (Jordan-H\"older theorem). Let 
$V_{mi}:{\cal H}_{mi}\rightarrow {\cal H}^\prime_{mi}$ be a
${\cal X}$-isomorphism. Then, $P_{m^\prime j}V_{mi}$ is a ${\cal
X}$-homomorphism from ${\cal H}_{mi}$ to
${\cal H}_{m^\prime j}$ and hence $P_{m^\prime j}V_{mi}=0$
if 
$m\neq m^\prime$ (Schur's lemma).
This implies that ${\cal H}^\prime_{mi}$ is a subspace of
 $\bigoplus_i{\cal H}_{mi}$. We thus have $P^\prime_mP_m=P^\prime_m$,
and similarly, $P^\prime_mP_m=P_m$, hence $P_m=P^\prime_m$.
Since ${\cal H}_{m1}$ and ${\cal H}^\prime_{m1}$ are 
${\cal X}$-isomorphic, 
${\rm Tr}(\rho_{\rm all}P_{m1})={\rm Tr}(\rho_{\rm all}P^\prime_{m1})$.
The algebra $\tilde{\cal X}$ and $\{\tilde\rho_s\}$ are thus uniquely
defined by Eq.~(\ref{def_tilde_rho}) when
$\{\rho_s\}$ and 
$\rho_{\rm all}$ are given.
 
Since 
${\cal H}_{\rm A}=\bigoplus_l(\bigoplus_k {\cal H}^{(l,k)})$
is also a decomposition of ${\cal H}_{\rm A}$ into 
simple ${\cal X}$-submodules, we can calculate 
$\tilde{\rho}_s$ as follows.
The form (\ref{rhodecompmax}) of $\rho_s$ assures that 
$\rho_{\rm all}$ is written as  
\begin{equation}
{\Gamma}\rho_{\rm all}{\Gamma}^\dagger
=\bigoplus_l {p}^{(l)}_{\rm all}\rho^{({\rm all},l)}_{\rm J} \otimes
\rho^{(l)}_{\rm K},
\label{decomprhoall}
\end{equation}
where $\rho^{({\rm all},l)}_{\rm J}$ 
are normalized
density operators acting on ${\cal H}^{(l)}_{\rm J}$.
Eq.~(\ref{Supprhoall}) assures that $p^{(l)}_{\rm
all}>0$. If we write the projection onto ${\cal H}^{(l,k)}$
as $P^{(l,k)}$, we have 
${\rm Tr}(\rho_{\rm
all}P^{(l,k)})={p}^{(l)}_{\rm all}q_k^{(l)}$.
Then we obtain 
\begin{equation}
{\Gamma}\tilde\rho_{s}{\Gamma}^\dagger
=\bigoplus_l \frac{{p}^{(s,l)}}{{p}^{(l)}_{\rm all}}\rho^{(s,l)}_{\rm J}
\otimes {\bf 1}^{(l)}_{\rm K}.
\end{equation}
It is now obvious that ${\cal H}^{(l,k)}$ and 
${\cal H}^{(l^\prime,k^\prime)}$ is 
$\tilde{\cal X}$-isomorphic when $l=l^\prime$.
It is also easy to show that they are not 
$\tilde{\cal X}$-isomorphic when $l\neq l^\prime$,
using a similar argument as above 
(Eq.~(\ref{isomorphic1}) changes to 
$ \Lambda^\prime(p^{(s,l)}/{p}^{(l)}_{\rm all})  \rho^{(s,l)}_{\rm J}
= (p^{(s,l^\prime)}/p^{(l^\prime)}_{\rm all}) \rho^{(s,l^\prime)}_{\rm
J} \Lambda^\prime$ in this case).

Using the property of the algebra $\tilde{\cal X}$,
we can prove the following lemma
\begin{lemma}
The maximal structure is unique. 
\label{maxisunique}
\end{lemma}
Let  $\Gamma:{\cal H}_{\rm A}\rightarrow
\bigoplus_l{\cal H}^{(l)}_{\rm J}\otimes
{\cal H}^{(l)}_{\rm K}$ be an isomorphism
corresponding to a maximal structure $D$,
and 
$\bar{\Gamma}:{\cal H}_{\rm A}\rightarrow
\bigoplus_l\bar{\cal H}^{(l)}_{\rm J}\otimes
\bar{\cal H}^{(l)}_{\rm K}$
be an isomorphism
corresponding to a maximal structure $\bar{D}$.
Take a basis $\{|b_k\rangle^{(l)}_{\rm K}\}$ for 
${\cal H}^{(l)}_{\rm K}$, and 
$\{|\bar{b}_k\rangle^{(l)}_{\rm K}\}$ for
 $\bar{\cal H}^{(l)}_{\rm K}$. 
Let ${\cal H}^{(l,k)}$ be the image of 
${\cal H}^{(l)}_{\rm J}\otimes|b_k\rangle^{(l)}_{\rm K}$
by $\Gamma^\dagger$, and $\bar{\cal H}^{(l,k)}$
be the image of $\bar{\cal H}^{(l)}_{\rm J}
\otimes|\bar{b}_k\rangle^{(l)}_{\rm
K}$ by $\bar\Gamma^\dagger$.
By appropriately
choosing the order of the index $l$, we can make
${\rm dim}\bar{\cal H}^{(l)}_{\rm K}=
{\rm dim}{\cal H}^{(l)}_{\rm K}$, and
${\cal H}^{(l,k)}$
be $\tilde{\cal X}$-isomorphic to
$\bar{\cal H}^{(l,k^\prime)}$   if and only
if 
$l=l^\prime$ (Jordan-H\"older theorem).
Through the isomorphisms $\Gamma$ and $\bar{\Gamma}$,
$\bigoplus_l{\cal H}^{(l)}_{\rm J}\otimes
{\cal H}^{(l)}_{\rm K}$ and 
$\bigoplus_l\bar{\cal H}^{(l)}_{\rm J}\otimes
\bar{\cal H}^{(l)}_{\rm K}$ can be regarded as 
$\tilde{\cal X}$-modules. Two $\tilde{\cal X}$-submodules
${\cal H}^{(l)}_{\rm J}\otimes|b_k\rangle^{(l)}_{\rm K}$
 and $\bar{\cal H}^{(l^\prime)}_{\rm J}
\otimes|\bar{b}_{k^\prime}\rangle^{(l^\prime)}_{\rm
K}$ are $\tilde{\cal X}$-isomorphic if and only
if $l=l^\prime$. Since $\bar\Gamma\Gamma^\dagger$ is 
a unitary $\tilde{\cal X}$-isomorphism, it is a direct sum 
of unitary $\tilde{\cal X}$-isomorphisms $V^{(l)}:{\cal H}^{(l)}_{\rm
J}\otimes {\cal H}^{(l)}_{\rm K}\rightarrow
\bar{\cal H}^{(l)}_{\rm J}\otimes
\bar{\cal H}^{(l)}_{\rm K}$ (Schur's lemma).
Note that $V^{(l)\dagger}(=(V^{(l)})^{-1})$ is 
also a $\tilde{\cal X}$-isomorphism.
Let $P^{(l)}_k$ be the projection onto 
${\cal H}^{(l)}_{\rm J}\otimes|b_k\rangle^{(l)}_{\rm K}$,
and $\bar{P}^{(l)}_k$ be the projection onto
$\bar{\cal H}^{(l)}_{\rm J}
\otimes|\bar{b}_{k}\rangle^{(l)}_{\rm
K}$. Without loss of generality, we assume that 
$\bar{P}^{(l)}_1V^{(l)}P^{(l)}_1\neq 0$.
Since $\bar{P}^{(l)}_1V^{(l)}P^{(l)}_1$
and $(\bar{P}^{(l)}_1V^{(l)}P^{(l)}_1)^\dagger$ are
 $\tilde{\cal X}$-homomorphisms, 
$(\bar{P}^{(l)}_1V^{(l)}P^{(l)}_1)^\dagger
(\bar{P}^{(l)}_1V^{(l)}P^{(l)}_1)=(c_{11})^2P^{(l)}_1$ with $c_{11}>0$
(Schur's lemma) and hence
$\bar{P}^{(l)}_1V^{(l)}P^{(l)}_1=c_{11}V^{(l)}_{\rm J}\otimes
|\bar{b}_{1}\rangle^{(l)}_{\rm K} \langle b_1|$,
where $V^{(l)}_{\rm J}$ is a unitary map from 
${\cal H}^{(l)}_{\rm J}$ to $\bar{\cal H}^{(l)}_{\rm J}$.
Since ${\bf 1}^{(l)}_{\rm J}\otimes
|b_{1}\rangle^{(l)}_{\rm K} \langle b_k|$ 
and 
 $\bar{\bf 1}^{(l)}_{\rm J}\otimes
|\bar{b}_{k^\prime}\rangle^{(l)}_{\rm K} \langle \bar{b}_k|$
are 
$\tilde{\cal X}$-isomorphisms, 
$V^{(l)}_{\rm J}\otimes
|\bar{b}_{k^\prime}\rangle^{(l)}_{\rm K} \langle b_k|$
is also a $\tilde{\cal X}$-isomorphism
 for any $k$ and $k^\prime$. Then, from
Schur's Lemma, 
$(V^{(l)}_{\rm J}\otimes
|\bar{b}_{k^\prime}\rangle^{(l)}_{\rm K} \langle b_k|)^\dagger
\bar{P}^{(l)}_{k^\prime}V^{(l)}P^{(l)}_k=c_{k^\prime k}P^{(l)}_k$
and we obtain 
$\bar{P}^{(l)}_{k^\prime}V^{(l)}P^{(l)}_k=c_{k^\prime k}
V^{(l)}_{\rm J}\otimes
|\bar{b}_{k^\prime}\rangle^{(l)}_{\rm K} \langle b_k|$.
We thus obtain 
\begin{equation}
\bar\Gamma\Gamma^\dagger=\bigoplus_l V^{(l)}
=\bigoplus_l\bigoplus_{k,k^\prime}
\bar{P}^{(l)}_{k^\prime}V^{(l)}P^{(l)}_k
=\bigoplus_l V^{(l)}_{\rm J}\otimes V^{(l)}_{\rm K},
\end{equation}
where 
$V^{(l)}_{\rm K}:{\cal H}^{(l)}_{\rm K}\rightarrow
\bar{\cal H}^{(l)}_{\rm K}$ is unitary since 
$\bar\Gamma\Gamma^\dagger$ and
$V^{(l)}_{\rm K}$ are unitary.
This means that the two structures, $D$ and $\bar{D}$,
are equivalent [see Eq.(\ref{Gammaunique})], and the lemma is proved.

Let us write this maximal structure 
as $D_{\max}(\{\rho_s\})$, which is uniquely determined when 
$\{\rho_s\}$ is given.
Lemma \ref{finismax} and Lemma \ref{maxisunique}
mean that the procedure described in Sec.~\ref{sec:main}
always yields a unique maximal structure. This also means that
if a structure is found to be maximal, it must satisfy the 
properties of $D_{\rm fin}$ derived in Sec.~\ref{sec:main}.
It will be convenient to state this in the form of a theorem:
\begin{theorem}
Let $\{\rho_s\}_{s\in S}$  be a set of density operators acting on 
${\cal H}_{\rm A}^\prime$. Suppose that the dimension 
of 
${\cal H}_{\rm A}\equiv \bigcup_{s\in S} {\rm Supp}(\rho_s)$
is finite.
Let $\Gamma:{\cal H}_{\rm A}\rightarrow
\bigoplus_l{\cal H}^{(l)}_{\rm J}\otimes
{\cal H}^{(l)}_{\rm K}$ be a unitary isomorphism  
that corresponds to a maximal structure $D_{\max}(\{\rho_s\}_{s\in S})$.
Then, any unitary operator $U$ acting on 
${\cal H}_{\rm A}^\prime\otimes {\cal
H}_{\rm E}$ that satisfies  ${\cal T}_U(\rho_s)=\rho_s$
for any $s\in S$ is expressed 
in the following
form under the isomorphism $\Gamma$,
\begin{equation}
U(\bbox{1}_{\rm A}\otimes \Sigma_{\rm E})
=\bigoplus_l\bbox{1}^{(l)}_{\rm J}\otimes U^{(l)}_{\rm
KE}(\bbox{1}^{(l)}_{\rm K}\otimes \Sigma_{\rm E}), 
\end{equation}
where $U^{(l)}_{\rm KE}$ are unitary operators acting on the combined
space ${\cal H}^{(l)}_{\rm K}\otimes {\cal H}_{\rm E}$.
\end{theorem}

Finally, let us consider the situation in which 
system A is made up of subsystems such that 
${\cal H}_{\rm A}= {\cal H}_{\rm A1}\otimes {\cal H}_{\rm
A2}\otimes\cdots$,
and the preparation of 
the initial state of system A is {\em independently} done for
each subsystem ${\cal H}_{{\rm A}i}$. 
In this case, the maximal
structure for ${\cal H}_{\rm A}$ is simply given by the 
`direct product' of the maximal structures for each subsystem,
as shown by the following theorem.

\begin{theorem}
Let $\{\rho_s\}_{s\in S_1}$ be density operators acting on 
${\cal H}_{\rm A1}$, and $\{\sigma_s\}_{s\in S_2}$ be
density operators acting on ${\cal H}_{\rm A2}$.
Suppose that the dimensions
of 
${\cal H}_{\rm A1}\equiv \bigcup_{s\in S_1} {\rm Supp}(\rho_s)$
and ${\cal H}_{\rm A2}\equiv \bigcup_{s\in S_2} {\rm Supp}(\sigma_s)$
is finite.
Let $\Gamma_1:{\cal H}_{\rm A1}\rightarrow
\bigoplus_{l_1}{\cal H}^{(l_1)}_{\rm J1}\otimes
{\cal H}^{(l_1)}_{\rm K1}$ be a unitary isomorphism  
that corresponds to a maximal structure $D_{\max}(\{\rho_s\}_{s\in S_1})$,
and $\Gamma_2:{\cal H}_{\rm A2}\rightarrow
\bigoplus_{l_2}{\cal H}^{(l_2)}_{\rm J2}\otimes
{\cal H}^{(l_2)}_{\rm K2}$ be a unitary isomorphism  
that corresponds to a maximal structure $D_{\max}(\{\sigma_s\}_{s\in
S_2})$. 
Define ${\cal H}_{\rm A}\equiv {\cal H}_{\rm A1}\otimes {\cal H}_{\rm
A2}$, ${\cal H}^{(l)}_{\rm J}\equiv 
{\cal H}^{(l_1)}_{\rm J1}\otimes {\cal H}^{(l_2)}_{\rm J2}$, and
${\cal H}^{(l)}_{\rm K}\equiv 
{\cal H}^{(l_1)}_{\rm K1}\otimes {\cal H}^{(l_2)}_{\rm K2}$,
where $l$ represents the double index $\{l_1,l_2\}$.
Then, $\Gamma\equiv \Gamma_1 \otimes \Gamma_2:
{\cal H}_{\rm A}\rightarrow
\bigoplus_l{\cal H}^{(l)}_{\rm J}\otimes
{\cal H}^{(l)}_{\rm K}$ corresponds to the maximal structure
$D_{\rm max}(\{\rho_s\otimes \sigma_t\}_{s\in S_1, t \in S_2})$.
\end{theorem}

This theorem implies that the collective operation 
to independently prepared systems has the same power
as individual operations.
For the proof, we will show that $\Gamma$ satisfies the three 
conditions of Definition 1.
From ${\Gamma_1}\rho_s{\Gamma}_1^\dagger
=\bigoplus_{l_1} {p}^{(s,l)}\rho^{(s,l_1)}_{\rm J1} \otimes
\rho^{(l_1)}_{\rm K1}$ and 
${\Gamma_2}\sigma_t{\Gamma}_2^\dagger
=\bigoplus_{l_2} {q}^{(t,l_2)}\sigma^{(t,l_2)}_{\rm J2} \otimes
\sigma^{(l_2)}_{\rm K2}$,
we have 
\begin{equation}
{\Gamma}\rho_s\otimes \sigma_t{\Gamma}^\dagger
=\bigoplus_{l_1,l_2}{p}^{(s,l_1)}{q}^{(t,l_2)}
(\rho^{(s,l_1)}_{\rm J1} \otimes \sigma^{(t,l_2)}_{\rm J2})
\otimes 
(\rho^{(l_1)}_{\rm K1}\otimes
\sigma^{(l_2)}_{\rm K2}),
\end{equation}
which means $\Gamma$ satisfies the condition (i) of Definition 1.
Next, 
construct a density operator $\sigma_{\rm all}$ by a
linear combination of the states 
$\{\sigma_t\}_{t\in S_2}$, such
that 
${\rm Supp}(\sigma_{\rm all})={\cal H}_{\rm A2}$ (see 
Sec.~\ref{formulation}). Then, $\sigma_{\rm all}$ is 
written as 
$
{\Gamma}_2\sigma_{\rm all}{\Gamma}_2^\dagger
=\bigoplus_{l_2} {q}^{(l_2)}_{\rm all}\sigma^{({\rm all},l_2)}_{\rm J2}
\otimes
\sigma^{(l_2)}_{\rm K2}$, where ${q}^{(l_2)}_{\rm all}>0$
[see Eq.~(\ref{decomprhoall})].
Take a basis $\{|k\rangle^{(l_2)}\}$ of
${\cal H}^{(l_2)}_{\rm J2}$
that diagonalizes $\sigma^{({\rm all},l_2)}_{\rm J2}$,
namely, $\sigma^{({\rm all},l_2)}_{\rm
J2}|k\rangle^{(l_2)}=c^{(l_2)}_k|k\rangle^{(l_2)}$ with 
$c^{(l_2)}_k>0$.
Suppose that for a value of $l=\{l_1,l_2\}$,
a projection operator 
$P:{\cal H}^{(l)}_{\rm J}\rightarrow {\cal H}^{(l)}_{\rm J}$
 satisfies 
$P{p}^{(s,l_1)}{q}^{(t,l_2)}(\rho^{(s,l_1)}_{\rm J1} \otimes
\sigma^{(t,l_2)}_{\rm J2}) ={p}^{(s,l_1)}{q}^{(t,l_2)}(\rho^{(s,l_1)}_{\rm
J1} \otimes
\sigma^{(t,l_2)}_{\rm J2})P$ for all $s$ and $t$.
Then, $\sigma^{({\rm all},l_2)}_{\rm J2}$ also satisfies 
\begin{equation}
P({p}^{(s,l_1)}\rho^{(s,l_1)}_{\rm J1} \otimes \sigma^{({\rm
all},l_2)}_{\rm J2}) =({p}^{(s,l_1)}\rho^{(s,l_1)}_{\rm J1} \otimes
\sigma^{({\rm all},l_2)}_{\rm J2})P
\label{prooflemma7_1}
\end{equation}
for all $s$.
$P$ can generally be  written as $P=\sum_{kk^\prime}
A_{k^\prime k}\otimes|k^\prime\rangle^{(l_2)}\langle k|$, where 
$A_{k^\prime k}$ are operators acting on 
${\cal H}^{(l_1)}_{\rm J1}$. Substituting it into
Eq.~(\ref{prooflemma7_1}), 
 we have $A_{k^\prime k}{p}^{(s,l_1)}\rho^{(s,l_1)}_{\rm J1}
=\beta {p}^{(s,l_1)}\rho^{(s,l_1)}_{\rm J1}A_{k^\prime k}$ for all
$s$, where $\beta= c^{(l_2)}_{k^\prime}
/ c^{(l_2)}_{k}$,
and hence $A_{k^\prime k}=\alpha_{k^\prime k}
{\bf 1}^{(l_1)}_{\rm J1}$ (Lemma \ref{Schur}).
$P$ is thus written as 
$P={\bf 1}^{(l_1)}_{\rm J1}\otimes B_{\rm J2}$, where
$B_{\rm J2}$ is an operator on ${\cal H}^{(l_2)}_{\rm J2}$.
A similar argument with ${\cal H}^{(l_1)}_{\rm J1}$ and 
${\cal H}^{(l_2)}_{\rm J2}$ interchanged leads to the form
$P=B_{\rm J1}\otimes{\bf 1}^{(l_2)}_{\rm J2}$, where
$B_{\rm J1}$ is an operator on ${\cal H}^{(l_1)}_{\rm J1}$.
Noting that $P$ is a projector, we conclude that
$P={\bf 1}^{(l_1)}_{\rm J1}\otimes{\bf 1}^{(l_2)}_{\rm J2}$
or $P={\bf 0}$, which means $\Gamma$ satisfies the condition (ii) of
Definition 1. Finally, suppose that, without loss of generality,
an operator 
$\Lambda:{\cal H}^{(1)}_{\rm J1}\otimes{\cal H}^{(l_2)}_{\rm J2}
\rightarrow 
{\cal H}^{(2)}_{\rm J1}\otimes{\cal H}^{(l^\prime_2)}_{\rm J2}$
 satisfies $p^{(s,1)}q^{(t,l_2)}
\Lambda(\rho^{(s,1)}_{\rm J1} \otimes \sigma^{(t,l_2)}_{\rm
J2}) =p^{(s,2)}q^{(t,l^\prime_2)}(\rho^{(s,2)}_{\rm J1} \otimes
\sigma^{(t,l^\prime_2)}_{\rm J2})\Lambda$ for all $s$ and $t$.
Then we have 
\begin{equation}
p^{(s,1)}q^{(l_2)}_{\rm all}\Lambda(\rho^{(s,1)}_{\rm J1} \otimes
\sigma^{({\rm all},l_2)}_{\rm J2}) =p^{(s,2)}q^{(l^\prime_2)}_{\rm all}
(\rho^{(s,2)}_{\rm
J1} \otimes
\sigma^{({\rm all},l^\prime_2)}_{\rm J2})\Lambda
\label{prooflemma7_2}
\end{equation}
for all $s$. 
$\Lambda$ can generally be  written as $\Lambda=\sum_{kk^\prime}
A_{k^\prime k}\otimes|k^\prime\rangle^{(l^\prime_2)}{}^{(l_2)}\langle k|$,
where 
$A_{k^\prime k}$ are operators from 
${\cal H}^{(1)}_{\rm J1}$ to ${\cal H}^{(2)}_{\rm J1}$. 
Substituting it
into Eq.~(\ref{prooflemma7_2}), 
 we have $A_{k^\prime k}p^{(s,1)}\rho^{(s,1)}_{\rm J1}
=\alpha p^{(s,2)}\rho^{(s,2)}_{\rm J1}A_{k^\prime k}$ for all
$s$, where 
$\alpha=q^{(l^\prime_2)}_{\rm all} c^{(l^\prime_2)}_{k^\prime}
/(q^{(l_2)}_{\rm all} c^{(l_2)}_{k})>0$.
From Lemma \ref{Schur}, we have $A_{k^\prime k}={\bf 0}$
and hence $\Lambda={\bf 0}$,
which means $\Gamma$ satisfies the condition (iii) of
Definition 1.

To summarize this section, we introduced a
structure called `maximal', that is uniquely 
defined when $\{\rho_s\}$ is given.
A set of conditions (see Definition 1) was given to 
check whether a given structure is maximal or not.
Given a maximal structure,  requirement for 
the operations to preserve $\{\rho_s\}$ is stated 
in a simple manner. The procedure described in 
Sec.~\ref{sec:main} gives a way to find a maximal 
structure in finite steps. Alternatively, a maximal 
structure is obtained by constructing the algebra 
$\tilde{\cal X}$ and by decomposing 
$\tilde{\cal X}$-module ${\cal H}_{\rm A}$ into 
simple $\tilde{\cal X}$-submodules, just like in 
finding irreducible representations for a group.

\section{Faithful transfer of quantum states}
\label{sec:transfer}

In the problem considered so far, 
the initial state of system A and the final state of 
the same physical system A are required to be identical.
In the problems concerning with communication, 
we often encounter a slightly different situation,
 in which the initial state of system A (held by the sender) 
and the final state 
of another physical system B (held by the receiver) 
are required to be identical. 
Here we will make a remark
that this problem of faithful transfer of quantum states is 
essentially the same as the problem considered in the preceding sections. 
The equivalence may be
self-evident when the dimension of system A and that of system B are the
same, if we note that we can freely transfer the state from
system A to system B or vice versa.
When the dimensions of the two systems are different,
there is a subtlety 
in this transfer and it will be worth while providing a detailed
argument here. The argument may also help clarifying the notations 
used in Sec.~\ref{sec:application}, which discusses examples of 
communication problems.

Let ${\cal H}_{\rm A}^\prime$ and ${\cal H}_{\rm B}^\prime$ be 
the Hilbert spaces for system A and B, respectively,
and ${\cal H}_{\rm E}$ be the Hilbert space for an
auxiliary system E. 
Initially, system A is secretly 
prepared in a state $\rho_s(s\in S)$.
Systems B and E are prepared in a standard state
$\Sigma_{\rm B}\equiv |u\rangle_{\rm B}\langle u|$ 
and $\Sigma_{\rm E}\equiv |u\rangle_{\rm E}\langle u|$, respectively.
In order to define the faithful transfer, 
we should assume a correspondence between the two physical 
systems A and B beforehand. This correspondence is 
given by a unitary map (isomorphism) $W_{\rm B:A}:{\cal H}_{\rm A}
\rightarrow {\cal H}_{\rm B}$, where ${\cal H}_{\rm B}$
is a subspace of ${\cal H}_{\rm B}^\prime$ with the same dimension
as ${\cal H}_{\rm A}\equiv \bigcup_{s\in S} {\rm Supp}(\rho_s)$. 
Any physical operation of the transfer can be described by 
a unitary operation $U_{\rm ABE}$
 acting on 
${\cal H}_{\rm A}^\prime\otimes {\cal H}_{\rm B}^\prime\otimes 
{\cal H}_{\rm E}$. Let $\sigma_s$ be the reduced state of 
system $B$ after the operation of $U_{\rm ABE}$. 
The requirement for the faithful transfer of $\{\rho_s\}$
 is that the relation $\sigma_s=W_{\rm B:A}\rho_s W_{\rm B:A}^\dagger$
should hold for any $s\in S$.
As before, the condition for this requirement will be
given as  a requirement for the isometry $\bar{U}_{\rm ABE}:
{\cal H}_{\rm A}\otimes |u\rangle_{\rm B}\otimes 
|u\rangle_{\rm E}\rightarrow 
{\cal H}_{\rm A}^\prime\otimes {\cal H}_{\rm B}^\prime\otimes 
{\cal H}_{\rm E}$, which is a restriction of $U_{\rm ABE}$.
The condition $\sigma_s=W_{\rm B:A}\rho_s W_{\rm B:A}^\dagger$
is explicitly written as
\begin{equation}
{\rm Tr}_{\rm AE}[\bar{U}_{\rm ABE}(\rho_s\otimes \Sigma_{\rm B}\otimes
\Sigma_{\rm E})
\bar{U}_{\rm ABE}^\dagger]=W_{\rm B:A}\rho_s W_{\rm
B:A}^\dagger.
\label{def_exact_transfer}
\end{equation} 
In this problem, there is 
no requirement for the final state
of  system A, and we can make it in an arbitrary state by 
applying a unitary operation on systems A and E.
We can thus impose an additional 
requirement that the final state of system A should be a standard state
$\Sigma_{\rm A}\equiv |u\rangle_{\rm A}\langle u|$, without loss of
generality. We thus assume that the image of $\bar{U}_{\rm ABE}$
is contained in $|u\rangle_{\rm A}\otimes {\cal H}_{\rm B}^\prime
\otimes {\cal H}_{\rm E}$.
Let us define $V_{\rm B:A}\equiv 
|u\rangle_{\rm A}(W_{B:A}) {}_{\rm B}\langle u|$, which is a
unitary map from 
${\cal H}_{\rm A}\otimes |u\rangle_{\rm B}$
to $|u\rangle_{\rm A}\otimes {\cal H}_{\rm B}$.
Since the operator $\bar{U}_{\rm ABE}(V_{\rm
B:A}^\dagger\otimes\Sigma_{\rm E})$
is an isometry from 
$|u\rangle_{\rm A}\otimes {\cal H}_{\rm B}
\otimes |u\rangle_{\rm E}$
to $|u\rangle_{\rm A}\otimes {\cal H}_{\rm B}^\prime
\otimes {\cal H}_{\rm E}$,
it can be written as 
\begin{equation}
\bar{U}_{\rm ABE}(V_{\rm B:A}^\dagger\otimes\Sigma_{\rm E})
=\Sigma_{\rm A}\otimes \bar{U}_{\rm BE},
\label{def_U_BE}
\end{equation}
where $\bar{U}_{\rm BE}$ is an isometry from 
${\cal H}_{\rm B}\otimes 
|u\rangle_{\rm E}$
to
${\cal H}_{\rm B}^\prime\otimes 
{\cal H}_{\rm E}$.
Note that the relation 
\begin{equation}
\rho_s\otimes \Sigma_{\rm B}=
V_{\rm B:A}^\dagger(\Sigma_{\rm A}\otimes W_{\rm B:A}\rho_s W_{\rm
B:A}^\dagger ) V_{\rm B:A}
\end{equation}
holds for any $\rho_s$. Substituting this into
Eq.~(\ref{def_exact_transfer}) and using Eq.~(\ref{def_U_BE}),
we have
\begin{equation}
{\rm Tr}_{\rm E}[\bar{U}_{\rm BE}( W_{\rm B:A}\rho_s W_{\rm
B:A}^\dagger\otimes
\Sigma_{\rm E})
\bar{U}_{\rm BE}^\dagger]=W_{\rm B:A}\rho_s W_{\rm
B:A}^\dagger.
\end{equation} 
This means that $\bar{U}_{\rm BE}$ preserves the set of states 
$\{W_{\rm B:A}\rho_s W_{\rm B:A}^\dagger\}$, and the main result of
Sec.~\ref{sec:main} or Theorem 3 can be applied. Noting that the
isomorphic  relation is defined between ${\cal H}_{\rm A}$ and 
${\cal H}_{\rm B}$, we can write the result as
\begin{equation}
\bar{U}_{\rm BE}=
W_{\rm B:A}
\left(
\bigoplus_l\bbox{1}^{(l)}_{\rm J}\otimes U^{(l)}_{\rm
KE}
\right)
W_{\rm B:A}^\dagger.
\end{equation}
Combined with Eq.~(\ref{def_U_BE}), we arrived at the following 
theorem.
\begin{theorem}
Let $\{\rho_s\}_{s\in S}$  be a set of 
density operators acting on 
${\cal H}_{\rm A}^\prime$. Suppose that the dimension 
of 
${\cal H}_{\rm A}\equiv \bigcup_{s\in S} {\rm Supp}(\rho_s)$
is finite.
Let $\Gamma:{\cal H}_{\rm A}\rightarrow
\bigoplus_l{\cal H}^{(l)}_{\rm J}\otimes
{\cal H}^{(l)}_{\rm K}$ be a unitary isomorphism  
that corresponds to a maximal structure $D_{\max}(\{\rho_s\}_{s\in S})$.
Let $W_{\rm B:A}:{\cal H}_{\rm A}
\rightarrow {\cal H}_{\rm B}$ be a unitary map,
where ${\cal H}_{\rm B}$ is a subspace of ${\cal H}_{\rm B}^\prime$.
Then, any isometry
$\bar{U}_{\rm ABE}:
{\cal H}_{\rm A}\otimes |u\rangle_{\rm B}\otimes 
|u\rangle_{\rm E}\rightarrow 
|u\rangle_{\rm A}\otimes {\cal H}_{\rm B}^\prime\otimes 
{\cal H}_{\rm E}$
that satisfies
\begin{equation}
{\rm Tr}_{\rm AE}[\bar{U}_{\rm ABE}(\rho_s\otimes \Sigma_{\rm B}\otimes
\Sigma_{\rm E})
\bar{U}_{\rm ABE}^\dagger]=W_{\rm B:A}\rho_s W_{\rm
B:A}^\dagger
\end{equation} 
for any $s\in S$ is expressed 
in the following
form under the isomorphism $\Gamma$,
\begin{equation}
\bar{U}_{\rm ABE}
=|u\rangle_{\rm A} (W_{\rm B:A})
\left(
\bigoplus_l\bbox{1}^{(l)}_{\rm J}\otimes U^{(l)}_{\rm
KE}
\right)
{}_{\rm B}\langle u|,
\end{equation}
where $U^{(l)}_{\rm KE}$ are isometries from 
${\cal H}^{(l)}_{\rm K}\otimes |u\rangle_{\rm E}$
to ${\cal H}^{(l)}_{\rm K}\otimes {\cal H}_{\rm E}$.
\end{theorem}

\section{Application to various problems}
\label{sec:application}
In this section,
we apply the derived properties of the operations preserving a
set of states to various problems such as 
cloning, cryptography, and data compression.

\subsection {Broadcasting of mixed states}

No-broadcasting condition
for  mixed states, which was derived in \cite{barnum96}, can easily
be rederived. The broadcasting is the task of preparing 
the marginal state of a subsystem of E in $\rho_s$, and leaving
the reduced state of the system A undisturbed as in $\rho_s$.
Since the operations that do not disturb $\{\rho_s\}$
are insensitive to the state changes in the subspaces 
${\cal H}^{(l)}_{\rm J}$, complete broadcasting is possible only when the 
dimensions of the subspaces ${\cal H}^{(l)}_{\rm J}$ are all unity, 
or equivalently, when $\{\rho_s\}$ can be simultaneously diagonalized.

In addition to rederiving this criteria, the derived principle here can
also determine the feasibility of various correlations between the two
broadcast systems, which was raised as an open question in \cite{barnum96}.
Let us consider the broadcasting of $\{\rho_s\}$ in system A into 
the two systems B and C. Let $H_{\rm X}^\prime ({\rm X=A,B,C})$ be 
the Hilbert space for system X, and 
suppose that the dimension 
of 
${\cal H}_{\rm A}\equiv \bigcup_{s\in S} {\rm Supp}(\rho_s)$
is finite. 
Take a subspace $H_{\rm X}\subset H_{\rm X}^\prime({\rm X=B,C})$
with the same size as ${\cal H}_{\rm A}$, and 
let $W_{\rm B:A}:{\cal H}_{\rm A}
\rightarrow {\cal H}_{\rm B}$ and 
$W_{\rm C:A}:{\cal H}_{\rm A}
\rightarrow {\cal H}_{\rm C}$ be unitary maps
defining the relation among the three systems.
 The process of broadcasting is defined as 
\begin{equation}
\rho_s \otimes \Sigma_{\rm B} \otimes \Sigma_{\rm C}
\rightarrow \Sigma_{\rm A}\otimes \chi^{(s)}_{\rm BC}
\label{phys_broad}
\end{equation}
with 
\begin{equation}
{\rm Tr}_{\rm C}(\chi^{(s)}_{\rm BC})=W_{\rm B:A}\rho_s W_{\rm
B:A}^\dagger, \;\;
{\rm Tr}_{\rm B}(\chi^{(s)}_{\rm BC})=W_{\rm C:A}\rho_s W_{\rm
C:A}^\dagger,
\label{def_broad}
\end{equation}
where  $\Sigma_{\rm X}\equiv |u\rangle_{\rm X} \langle u|({\rm X=A,B,C})$
are standard states.
When the broadcasting is possible, 
the supporting space ${\cal H}_{\rm A}$ can 
be decomposed as ${\cal H}_{\rm A}=\bigoplus_l {\cal H}^{(l)}_{\rm K}$,
since 
${\cal H}^{(l)}_{\rm J}$ is one-dimensional and can thus be neglected. 
Then, by taking appropriate bases $\{|a_k\rangle^{(l)}_{\rm K}\}$
for ${\cal H}^{(l)}_{\rm K}$, we can write
$\rho_s=\bigoplus_l p^{(s,l)} \rho^{(l)}_{\rm K}=
\sum_l\sum_k p^{(s,l)}q^{(l)}_k|a_k\rangle^{(l)}_{\rm
K}
\langle a_k|$. 
Let us take bases for ${\cal H}_{\rm B}$ and ${\cal H}_{\rm C}$
by $|a_k\rangle^{(l)}_{\rm B}\equiv W_{\rm B:A}|a_k\rangle^{(l)}_{\rm K}$
and $|a_k\rangle^{(l)}_{\rm C}\equiv W_{\rm C:A}|a_k\rangle^{(l)}_{\rm
K}$. The broadcast state $\chi^{(s)}_{\rm BC}$ satisfying 
Eq.~(\ref{def_broad}) is not unique and various types of correlations
between systems B and C are conceivable. For example, 
a state with no correlation
\begin{equation}
\chi^{(s)}_{\rm BC}=W_{\rm B:A}\rho_s W_{\rm
B:A}^\dagger\otimes
W_{\rm C:A}\rho_s W_{\rm
C:A}^\dagger,
\label{no_corr_broad}
\end{equation}
a state with classical correlations
\begin{equation}
\chi^{(s)}_{\rm BC}=\sum_l\sum_k p^{(s,l)}q^{(l)}_k
|a_k\rangle^{(l)}_{\rm B}
\langle a_k|\otimes |a_k\rangle^{(l)}_{\rm C}
\langle a_k|,
\end{equation}
and a state with quantum correlations 
$\chi^{(s)}_{\rm BC}=|\chi^{(s)}\rangle\langle \chi^{(s)}|$ with 
\begin{equation}
|\chi^{(s)}\rangle=\sum_l\sum_k 
\exp(i\theta_{l,k})
\sqrt{p^{(s,l)}q^{(l)}_k}
|a_k\rangle^{(l)}_{\rm B}
|a_k\rangle^{(l)}_{\rm C}
\label{qu_corr_broad}
\end{equation}
all satisfies Eq.~(\ref{def_broad}). The question here is,
among these and other conceivable correlations, what are feasible 
by a physical process Eq.~(\ref{phys_broad}).
To answer this problem, let us start by noting that 
any physical process acting on $\{\rho_s\}$ corresponds to an isometry
$\bar{U}_{\rm ABCE}:
{\cal H}_{\rm A}\otimes |u\rangle_{\rm B}\otimes
|u\rangle_{\rm C}\otimes 
|u\rangle_{\rm E}\rightarrow 
|u\rangle_{\rm A}\otimes {\cal H}_{\rm B}^\prime 
\otimes {\cal H}_{\rm
C}^\prime\otimes  {\cal H}_{\rm E}$
with an auxiliary system E.
Along with the decomposition ${\cal H}_{\rm A}=\bigoplus_l {\cal
H}^{(l)}_{\rm K}$, we can decompose 
$\bar{U}_{\rm ABCE}$ as 
 $\bar{U}_{\rm ABCE}=\bigoplus_l \bar{U}_{\rm ABCE}^{(l)}$
by isometries $\bar{U}_{\rm ABCE}^{(l)}:
{\cal H}^{(l)}_{\rm K}\otimes |u\rangle_{\rm B}\otimes
|u\rangle_{\rm C}\otimes 
|u\rangle_{\rm E}\rightarrow 
|u\rangle_{\rm A}\otimes {\cal H}_{\rm B}^\prime 
\otimes {\cal H}_{\rm
C}^\prime\otimes  {\cal H}_{\rm E}$.
Since $\rho_s$ is exactly transferred from A to B, we
have, from  Theorem 5,
\begin{equation}
\bar{U}_{\rm ABCE}
=|u\rangle_{\rm A} (W_{\rm B:A})
\left(
\bigoplus_l\bbox{1}^{(l)}_{\rm J}\otimes U^{(l)}_{\rm
KCE}
\right)
{}_{\rm B}\langle u|,
\end{equation}
where $U^{(l)}_{\rm KCE}$ are isometries from 
${\cal H}^{(l)}_{\rm K}\otimes |u\rangle_{\rm C}\otimes |u\rangle_{\rm E}$
to ${\cal H}^{(l)}_{\rm K}\otimes 
{\cal H}_{\rm C}^\prime\otimes {\cal H}_{\rm
E}$.
Noting that we are omitting ${\cal H}^{(l)}_{\rm J}$, we have 
\begin{equation}
\bar{U}^{(l)}_{\rm ABCE}
=|u\rangle_{\rm A} (W_{\rm B:A})
(U^{(l)}_{\rm
KCE})
{}_{\rm B}\langle u|.
\end{equation}
This means that the image of $\bar{U}^{(l)}_{\rm ABCE}$ is
contained in $|u\rangle_{\rm A}\otimes {\cal H}^{(l)}_{\rm B} 
\otimes {\cal H}_{\rm
C}^\prime\otimes  {\cal H}_{\rm E}$,
where ${\cal H}^{(l)}_{\rm B}$ is the image of 
${\cal H}^{(l)}_{\rm K}$ by $W_{\rm B:A}$.
Similarly, since
 $\rho_s$ is exactly transferred from A to C, we have
another expression,
\begin{equation}
\bar{U}^{(l)}_{\rm ABCE}
=|u\rangle_{\rm A} (W_{\rm C:A})
(U^{(l)}_{\rm
KBE})
{}_{\rm C}\langle u|,
\end{equation}
where $U^{(l)}_{\rm KBE}$ are isometries from 
${\cal H}^{(l)}_{\rm K}\otimes |u\rangle_{\rm B}\otimes |u\rangle_{\rm E}$
to ${\cal H}^{(l)}_{\rm K}\otimes 
{\cal H}_{\rm B}^\prime\otimes {\cal H}_{\rm
E}$.
By this expression, the image of $\bar{U}^{(l)}_{\rm ABCE}$ is
further restricted to $|u\rangle_{\rm A}\otimes {\cal H}^{(l)}_{\rm B} 
\otimes {\cal H}^{(l)}_{\rm C} \otimes  {\cal H}_{\rm E}$,
where ${\cal H}^{(l)}_{\rm C}$ is the image of 
${\cal H}^{(l)}_{\rm K}$ by $W_{\rm C:A}$.
The operation $\bar{U}_{\rm ABCE}$ thus only connects 
the subspaces labeled by the same value of index $l$.
The broadcast state $\chi^{(s)}_{\rm BC}$, which is given by
\begin{equation}
\chi^{(s)}_{\rm BC}={\rm Tr}_{\rm AE}
[\bar{U}_{\rm ABCE}(
\rho_s \otimes \Sigma_{\rm B} \otimes \Sigma_{\rm C}
\otimes \Sigma_{\rm E})\bar{U}_{\rm ABCE}^\dagger],
\end{equation}
should therefore be written as
\begin{equation}
\chi^{(s)}_{\rm BC}=\bigoplus_l p^{(s,l)}\zeta^{(l)}_{\rm BC}
\label{allowed_broad}
\end{equation}
where $\zeta^{(l)}_{\rm BC}$ is a density operator 
 acting on ${\cal H}^{(l)}_{\rm B} 
\otimes {\cal H}^{(l)}_{\rm C}$, given by
\begin{equation}
\zeta^{(l)}_{\rm BC}\equiv {\rm Tr}_{\rm AE}
[\bar{U}^{(l)}_{\rm ABCE}(
\rho^{(l)}_{\rm K} \otimes \Sigma_{\rm B} \otimes \Sigma_{\rm C}
\otimes \Sigma_{\rm E})\bar{U}_{\rm ABCE}^{(l)\dagger}].
\end{equation}
The condition (\ref{def_broad}) for broadcasting is satisfied iff
\begin{equation}
{\rm Tr}_{\rm C}(\zeta^{(l)}_{\rm BC})=W_{\rm B:A}\rho^{(l)}_{\rm K}
W_{\rm B:A}^\dagger, \;\;
{\rm Tr}_{\rm B}(\zeta^{(l)}_{\rm BC})=W_{\rm C:A}\rho^{(l)}_{\rm K}
W_{\rm C:A}^\dagger
\label{def_broad_l}
\end{equation}
holds for all $l$.
Since $\zeta^{(l)}_{\rm BC}$ is independent of $s$, it can be
any state by choosing $\bar{U}^{(l)}_{\rm ABCE}$ appropriately.
This means that any types of correlations are 
feasible in each subspace ${\cal H}^{(l)}_{\rm B} 
\otimes {\cal H}^{(l)}_{\rm C}$, ranging from
 quantum correlation (entanglement) to no correlation. 
On the other hand, Eq.~(\ref{allowed_broad}) means that
for the index $l$, a complete classical correlation 
should always be established between the broadcast systems. 

One of the interesting consequences from the above general result 
is that
the condition for the feasibility of the broadcast state with no
correlation  [Eq.(\ref{no_corr_broad})] and that of 
broadcast states with full quantum correlation 
[Eq.(\ref{qu_corr_broad})] are the same.
For both cases, the condition is that any $\rho_s$ should 
be contained in one of the subspaces ${\cal H}^{(l)}_{\rm K}$,
or equivalently, any pair of states from $\{\rho_s\}$ must be 
identical or orthogonal.

\subsection{Imprinting of mixed states}

Another open question was
the condition for the feasibility of the imprinting process\cite{mor98}.
The no-imprinting condition is the requirement for $\{\rho_s\}$ such 
that any attempt to read out the information on $s$ should lead to
some changes in the state of system A from the initial state. 
More formally, under the  notations used here, it is the condition for
$\{\rho_s\}$ such that for any unitary operator $U$ acting on 
${\cal H}_{\rm A}^\prime\otimes {\cal
H}_{\rm E}$ satisfying ${\rm Tr}_{\rm E}[U(\rho_s\otimes \Sigma_{\rm
E})U^\dagger]=\rho_s$, the reduced state of system E, 
${\rm Tr}_{\rm A}[U(\rho_s\otimes \Sigma_{\rm
E})U^\dagger]$, should be independent of $s$.

This condition is obvious now. According to 
the present result, such an operation $U$ is insensitive to 
the contents of ${\cal H}^{(l)}_{\rm J}$, and 
${\cal H}^{(l)}_{\rm K}$ holds no information on $s$.
On the other hand, the index $l$ can be read out freely 
without disturbing $\{\rho_s\}$.
Hence the condition is stated as 
 $p^{(s,l)}=p^{(s^\prime,l)}$
for all $\{s,s^\prime,l\}$, namely, the probability distribution for 
the index $l$ is identical for all $s$. In other words,
this is the requirement that if $\{\rho_s\}$ are written as matrices in 
the maximally simultaneously 
block-diagonalized form, the traces for each block 
are the same for all $s$. 

A generalized version of this theorem, where 
the set of states $\{\sigma_{s'}\}_{s'\in S'} $ to be 
distinguished is different from the set of states $\{\rho_s\}$
to be preserved, can also be derived from the above results.
A little care should be taken for the fact that 
the support of all states,
defined as
$\bar{\cal H}_{\rm A}=[ \bigcup_{s\in S} {\rm Supp}(\rho_s)]
\cup [\bigcup_{s'\in S'} {\rm Supp}(\sigma_{s'})]$, may generally be
larger than ${\cal H}_{\rm A}= \bigcup_{s\in S} {\rm Supp}(\rho_s)$.
Let us write $\bar{\cal H}_{\rm A}={\cal H}_{\rm A}\oplus 
{\cal H}_{0}$. 
Consider the set of all isometries $U:\bar{\cal H}_{\rm A}\otimes
|u\rangle_{\rm E}\rightarrow \bar{\cal H}_{\rm A}\otimes
{\cal H}_{\rm E}$ that preserve $\{\rho_s\}$, namely,
${\rm Tr}_{\rm E}[U(\rho_s\otimes \Sigma_{\rm
E})U^\dagger]=\rho_s$. What we ask here is the condition for 
$\{\sigma_{s'}\}_{s'\in S'} $ such that 
${\rm Tr}_{\rm A}[U(\sigma_{s'}\otimes \Sigma_{\rm
E})U^\dagger]$ is independent of $s'\in S'$ under any such $U$.
We first derive a sufficient condition. According to 
Theorem 3, a decomposition ${\cal H}_{\rm A}=
\bigoplus_l{\cal H}^{(l)}_{\rm J}\otimes {\cal H}^{(l)}_{\rm
K}$ is determined 
from $\{\rho_s\}$, and 
$U$ is written as 
\begin{equation}
U=U_{\rm 0E}\oplus 
\left(
\bigoplus_l\bbox{1}^{(l)}_{\rm J}\otimes U^{(l)}_{\rm
KE}\right),
\end{equation}
where $U^{(l)}_{\rm KE}$ are isometries from 
${\cal H}^{(l)}_{\rm K}\otimes |u\rangle_{\rm E}$
to ${\cal H}^{(l)}_{\rm K}\otimes {\cal H}_{\rm
E}$, and $U_{\rm 0E}$ is an isometry from 
${\cal H}_{0}\otimes |u\rangle_{\rm E}$
to $\bar{\cal H}_{\rm A}\otimes {\cal H}_{\rm
E}$. Note that the image of $U_{\rm 0E}$ is not necessarily
confined in ${\cal H}_0\otimes {\cal H}_{\rm
E}$.
Then we can write
\begin{eqnarray}
&&{\rm Tr}_{\rm A}[U(\sigma_{s'}\otimes \Sigma_{\rm
E})U^\dagger]={\rm Tr}_{\rm A}[U_{\rm 0E}(\sigma_{s'}\otimes
\Sigma_{\rm E})U_{\rm 0E}^\dagger]
\nonumber \\
&&+\sum_l\left(
{\rm Tr}_{\rm A}[U_{\rm 0E}(\sigma_{s'}\otimes
\Sigma_{\rm E})(\bbox{1}^{(l)}_{\rm J}\otimes U^{(l)}_{\rm
KE})^\dagger ]
\right.
\nonumber \\
&&+{\rm Tr}_{\rm A}[(\bbox{1}^{(l)}_{\rm J}\otimes U^{(l)}_{\rm
KE})(\sigma_{s'}\otimes
\Sigma_{\rm E})U_{\rm 0E}^\dagger ]
\nonumber \\
&&\left.
+{\rm Tr}_{\rm A}[(\bbox{1}^{(l)}_{\rm J}\otimes U^{(l)}_{\rm
KE})(\sigma_{s'}\otimes
\Sigma_{\rm E})(\bbox{1}^{(l)}_{\rm J}\otimes U^{(l)}_{\rm
KE})^\dagger ]
\right).
\label{state_E_imp}
\end{eqnarray}
Let $P_0$, $P_{\rm A}$,  and $P^{(l)}_{\rm A}$ be the projection 
operators onto ${\cal H}_{0}$, ${\cal H}_{\rm A}$, and 
${\cal H}^{(l)}_{\rm J}\otimes {\cal H}^{(l)}_{\rm
K}$, respectively. Equation (\ref{state_E_imp}) means that 
the state left in system E  depends only on the following 
parts of the initial state $\sigma_{s'}$, defined as 
\begin{eqnarray}
\sigma_{s'}^{(00)}&\equiv & P_0\sigma_{s'}P_0,
\nonumber \\
\sigma_{s'}^{(0{\rm A})}&\equiv  &P_0\sigma_{s'}P_{\rm A},
\;\; \sigma_{s'}^{({\rm A}0)}\equiv  P_{\rm A}\sigma_{s'}P_0,
\nonumber \\
\sigma_{s'}^{(l)}&\equiv  &
{\rm Tr}_{\rm J}^{(l)}(P^{(l)}_{\rm A}\sigma_{s'}P^{(l)}_{\rm A}),
\label{def_part_sigma}
\end{eqnarray}
and Eq.~(\ref{state_E_imp}) becomes 
\begin{eqnarray}
&&{\rm Tr}_{\rm A}[U(\sigma_{s'}\otimes \Sigma_{\rm
E})U^\dagger]={\rm Tr}_{\rm A}[U_{\rm 0E}(\sigma^{(00)}_{s'}\otimes
\Sigma_{\rm E})U_{\rm 0E}^\dagger]
\nonumber \\
&&+\sum_l\left(
{\rm Tr}_{\rm A}[U_{\rm 0E}(\sigma_{s'}^{(0{\rm A})}\otimes
\Sigma_{\rm E})(\bbox{1}^{(l)}_{\rm J}\otimes U^{(l)}_{\rm
KE})^\dagger ]
\right.
\nonumber \\
&&+{\rm Tr}_{\rm A}[(\bbox{1}^{(l)}_{\rm J}\otimes U^{(l)}_{\rm
KE})(\sigma_{s'}^{({\rm A}0)}\otimes
\Sigma_{\rm E})U_{\rm 0E}^\dagger ]
\nonumber \\
&&\left.
+{\rm Tr}_{\rm K}^{(l)}[ U^{(l)}_{\rm
KE}(\sigma_{s'}^{(l)}\otimes
\Sigma_{\rm E}) (U^{(l)}_{\rm
KE})^\dagger ]
\right). 
\end{eqnarray}
Here ${\rm Tr}_{\rm J}^{(l)}$ and ${\rm Tr}_{\rm K}^{(l)}$
are the partial trace over ${\cal H}^{(l)}_{\rm
J}$ and ${\cal H}^{(l)}_{\rm K}$, respectively.
Hence a sufficient condition for the state left in system 
E to have no dependence on $s'$ is that 
the operators defined in Eq.~(\ref{def_part_sigma}) are 
independent of $s'$.

To show that this condition is also necessary, we will consider
a particular measurement strategy. Let $l^*$ be one of the 
possible values of the index $l$. The first step of the
strategy is 
to conduct an ideal projection measurement to measure 
whether the state is in the subspace 
${\cal H}_0\oplus {\cal H}^{(l^*)}_{\rm J}\otimes 
{\cal H}^{(l^*)}_{\rm K}$ or not. It is obvious that this measurement 
does not disturb $\{\rho_s\}$. 
Let $p(l^*,s')$ be the probability of obtaining
the positive outcome when the initial state was $\sigma_{s'}$.
Suppose that the result of 
the measurement was positive. Prepare auxiliary physical 
systems J and K with Hilbert spaces ${\cal H}_{\rm J}$ and
${\cal H}_{\rm K}={\cal H}_{\rm K0}\oplus {\cal H}_{\rm K1}$,
respectively, where 
${\rm dim}\; {\cal H}_{\rm J}=
{\rm dim}\; {\cal H}^{(l^*)}_{\rm J}$,
${\rm dim}\; {\cal H}_{\rm K1}=
{\rm dim}\; {\cal H}^{(l^*)}_{\rm K}$,
and 
${\rm dim}\; {\cal H}_{\rm K0}=
{\rm dim}\; {\cal H}_0$. 
Take unitary maps $\Gamma_{\rm J}:
{\cal H}^{(l^*)}_{\rm J}\rightarrow
{\cal H}_{\rm J}$, 
$\Gamma_{\rm K}:
{\cal H}^{(l^*)}_{\rm K}\rightarrow
{\cal H}_{\rm K1}$, and 
$\Gamma_{0}:
{\cal H}_{0}\rightarrow
{\cal H}_{\rm K0}$. 
Take an arbitrary state $|x\rangle\in {\cal H}^{(l^*)}_{\rm J}$,
and let $\Gamma_{0}^\prime:
{\cal H}_{0}\rightarrow
(\Gamma_{\rm J}|x\rangle)\otimes {\cal H}_{\rm K0}$ be the unitary map 
naturally determined from $\Gamma_{0}$.
Then, we can construct an isometry $\Gamma:
{\cal H}_0\oplus {\cal
H}^{(l^*)}_{\rm J}\otimes  {\cal H}^{(l^*)}_{\rm K}
\rightarrow {\cal H}_{\rm
J}\otimes {\cal H}_{\rm K}$ by $\Gamma\equiv \Gamma_{0}^\prime
\oplus \Gamma_{\rm J}\otimes \Gamma_{\rm
K}$. The second step is to transfer the
post-measurement state of system A, which is projected in 
${\cal H}_0\oplus {\cal
H}^{(l^*)}_{\rm J}\otimes  {\cal H}^{(l^*)}_{\rm K}$, 
to the combined system of J and K according to the isometry
$\Gamma$. Note that if the initial state was $\rho_s$,
the state of ${\cal H}_{\rm
J}\otimes {\cal H}_{\rm K}$ after the second step is 
$\Gamma_{\rm J} \rho^{(s,l^*)}_{\rm J} \Gamma_{\rm J}^\dagger
\otimes \Gamma_{\rm K} \rho^{(l^*)}_{\rm K} \Gamma_{\rm K}^\dagger$.
The third step is to conduct arbitrary measurement on system K,
and to leave system K in  
$\Gamma_{\rm K} \rho^{(l^*)}_{\rm K} \Gamma_{\rm K}^\dagger$,
which is independent of the initial state. At the final step,
the state of ${\cal H}_{\rm
J}\otimes {\cal H}_{\rm K}$, which should be contained in 
the image of $\Gamma$, is transferred back to 
${\cal H}_0\oplus {\cal
H}^{(l^*)}_{\rm J}\otimes  {\cal H}^{(l^*)}_{\rm K}$.
It is easy to see that the whole process does not disturb 
$\{\rho_s\}$. When the initial state was $\sigma_{s'}$,
the marginal state of system K after the second step,
multiplied by $p(l^*,s')$,  is
\begin{eqnarray}
&&{\rm Tr}_{\rm J}[\Gamma
(P_0\oplus \bbox{1}^{(l^*)}_{\rm J}\otimes
\bbox{1}^{(l^*)}_{\rm K})
\sigma_{s'}(P_0\oplus \bbox{1}^{(l^*)}_{\rm J}\otimes
\bbox{1}^{(l^*)}_{\rm K})\Gamma^\dagger]
\nonumber \\
&=&\Gamma_0\sigma_{s'}^{(00)}\Gamma_0^\dagger
+\Gamma_0\sigma_{s'}^{(0{\rm A})}
(|x\rangle\otimes
\bbox{1}^{(l^*)}_{\rm K})\Gamma_{\rm K}^\dagger
\nonumber \\
&&+\Gamma_{\rm K}(\langle x| \otimes
\bbox{1}^{(l^*)}_{\rm K})\sigma_{s'}^{({\rm A}0)}
\Gamma_0^\dagger
+\Gamma_{\rm K}\sigma_{s'}^{(l^*)}
\Gamma_{\rm K}^\dagger,
\label{state_K_imp}
\end{eqnarray}
If we require, for this particular strategy, that
 the statistics of the outcomes
of the measurement of the first step and 
the arbitrary measurement in the third step 
should be independent of $s'$, the reduced state 
in Eq.~(\ref{state_K_imp}) must be independent 
of $s'$. Since the choices of $l^*$ and $|x\rangle$ 
were arbitrary, we conclude that it is necessary
that $\sigma_{s'}^{(00)}$, $\sigma_{s'}^{(0{\rm A})}$,
$\sigma_{s'}^{({\rm A}0)}$, and $\sigma_{s'}^{(l)}$ for any $l$
be independent of $s'$. 
It is worth emphasizing here that the above result shows that
 any difference in the off-diagonal part 
($\sigma_{s'}^{(0{\rm A})}$) between ${\cal H}_0$ and 
${\cal H}_{\rm A}$ is detectable even under a stringent 
restriction to the operation to ${\cal H}_{\rm A}$.

\subsection{Cloning and imprinting of composite systems}

Consider
the situation in which the system holding an unknown initial state
$\chi_s$ is composed of two subsystems A and B, and it is allowed to 
access these subsystems only in sequence, namely, subsystem
A must be released before subsystem 
B is accessed\cite{koashi98nov,mor98}. 
In order to preserve 
the states $\{\chi_s\}$ in
the whole system, the marginal
density operator in 
A, $\rho_s\equiv \mbox{Tr}_{\rm B}(\chi_s)$, 
must not be modified when it
is released. The present results can thus be applied and restrict
the form of the operation when subsystem A is at hand. 
Let us write this operation by an isometry 
$U_{\rm AE}:{\cal H}_{\rm A}\otimes
|u\rangle_{\rm E}\rightarrow 
{\cal H}_{\rm A}\otimes
{\cal H}_{\rm E}$, where
${\cal H}_{\rm A}= \bigcup_{s\in S} {\rm Supp}(\rho_s)$,
and ${\cal H}_{\rm E}$ describes an auxiliary system initially
prepared in a standard state 
$\Sigma_{\rm E}=|u\rangle_{\rm E}\langle u|$.
Let ${\cal H}_{\rm B}$  be the Hilbert space for subsystem B.
According to 
Theorem 3, a decomposition ${\cal H}_{\rm A}=
\bigoplus_l{\cal H}^{(l)}_{\rm J}\otimes {\cal H}^{(l)}_{\rm
K}$ is determined 
from $\{\rho_s\}$, and 
$U_{\rm AE}$ is written as 
\begin{equation}
U_{\rm AE}=
\bigoplus_l\bbox{1}^{(l)}_{\rm J}\otimes U^{(l)}_{\rm
KE},
\end{equation}
where $U^{(l)}_{\rm KE}$ are isometries from 
${\cal H}^{(l)}_{\rm K}\otimes |u\rangle_{\rm E}$
to ${\cal H}^{(l)}_{\rm K}\otimes {\cal H}_{\rm
E}$.
Then we can write
the marginal state $\chi^{(s)}_{\rm BE}$ 
of the combined system of E and B after the operation 
$U_{\rm AE}$ as
\begin{eqnarray}
\chi^{(s)}_{\rm BE}&\equiv&
{\rm Tr}_{\rm A}[(U_{\rm AE}\otimes
\bbox{1}_{\rm B})(\chi_{s}\otimes \Sigma_{\rm
E})(U_{\rm AE}^\dagger\otimes
\bbox{1}_{\rm B})]
\nonumber \\
&=&\sum_l
{\rm Tr}_{\rm A}[(\bbox{1}^{(l)}_{\rm J}\otimes U^{(l)}_{\rm
KE}\otimes
\bbox{1}_{\rm B})(\chi_{s}\otimes
\Sigma_{\rm E})(\bbox{1}^{(l)}_{\rm J}\otimes U^{(l)\dagger}_{\rm
KE}\otimes
\bbox{1}_{\rm B}) ]
\nonumber \\
&=&\sum_l
{\rm Tr}_{\rm K}^{(l)}[(U^{(l)}_{\rm
KE}\otimes
\bbox{1}_{\rm B})(\chi_{s}^{(l)}\otimes
\Sigma_{\rm E})( U^{(l)\dagger}_{\rm
KE}\otimes
\bbox{1}_{\rm B}) ].
\end{eqnarray}
where
\begin{equation}
\chi_{s}^{(l)}\equiv
{\rm Tr}_{\rm J}^{(l)}[
(P_{\rm A}^{(l)}\otimes
\bbox{1}_{\rm B})\chi_s
(P_{\rm A}^{(l)}\otimes
\bbox{1}_{\rm B})
].
\end{equation}
This means that if we are to preserve $\mbox{Tr}_{\rm B}(\chi_s)$,
we
can obtain only the part of the correlations between A and B, namely,
we can obtain classical correlations related to the index $l$ and 
quantum correlations related to each ${\cal H}^{(l)}_{\rm K}$,
but cannot obtain quantum correlations related to the index $l$ or 
any correlations related to each ${\cal H}^{(l)}_{\rm J}$. Note that
extracting this information  to the auxiliary system E
 may destroy original quantum
correlations between A and B.  

When $\{\chi_s\}$ are all
different pure states, the cloning of $\{\chi_s\}$
is possible only when $\{\chi_s\}$ are all orthogonal.
Hence, if the cloning is possible, we should be 
able to determine $s$ completely by conducting a measurement  
on $\chi^{(s)}_{\rm BE}$, namely,
$\{\chi^{(s)}_{\rm BE}\}$ should be all orthogonal.
This is possible only when
$\chi_{s}^{(l)}$ and $\chi_{s'}^{(l)}$
are orthogonal for any $l$ and for any $s\neq s'$.
This condition is also sufficient for the cloning to be possible,
since under that condition we can determine the value of $s$
by accessing B and E without disturbing the whole state 
$(U_{\rm AE}\otimes
\bbox{1}_{\rm B})(\chi_{s}\otimes \Sigma_{\rm
E})(U_{\rm AE}^\dagger\otimes
\bbox{1}_{\rm B})$.
At this stage,
we know exactly the current pure state of the system ABE,
and the marginal state of subsystem A 
is $\mbox{Tr}_{\rm B}(\chi_s)$. 
Then we can determine a unitary operation over 
BE that drives the state of ABE back into 
$\chi_{s}\otimes \Sigma_{\rm E}$.

When the initial states $\{\chi_s\}$ include mixed states,
it is not always possible to restore
the original quantum correlations between A and B 
 by manipulating system BE only. 
Also in the problem of imprinting of more than two pure states,
the restoration is not always possible because 
the identifying $s$ by accessing system BE
 is not necessarily possible. The
feasibility of this restoration of quantum correlation will be an
interesting future problem.

\subsection{Various schemes of quantum key distribution}
An attempt 
was recently made by Mor\cite{mor98}, 
to give a unified explanation of why various schemes of quantum key
distribution work in the ideal situations, which was based on 
the no-cloning principle of mixed states. Since we have obtained 
a general principle including the no-cloning principle,
we provide a unified formalism  of various schemes for quantum key
distribution.  The principle here places some
restriction on the eavesdropper's  access to the first quantum system A
transmitted from the sender to  the receiver, if 
the eavesdropper wants to preserve
the state in order to  conceal her presence. Then we can find three
different ways  to conceal the bit value from the eavesdropper, namely, 
(i) encoding it directly on the inner degree of freedom for ${\cal
H}_{\rm J}^{(l)}$, 
(ii) encoding it on the correlation between A and another system 
B, through the inner degree of freedom for ${\cal H}_{\rm J}^{(l)}$, 
and
(iii) encoding it on the {\it quantum} correlation between A and 
B, through the index $l$. The original 
four-state scheme of Bennett and Brassard\cite{bennett84} corresponds
to the case (ii), since the bit value is encoded on the correlation 
between the quantum state of the photon and the information of the 
basis transmitted later\cite{peres}, which corresponds to the system B.
The scheme \cite{bennett92} using two nonorthogonal pure states 
corresponds to the case (i), and the schemes using three\cite{goldenberg95}
or two\cite{koashi97} entangled states in the composite system correspond
to the case (iii).

\subsection{Optimal compression rate of quantum-state signals}

Consider a source that produces the ensemble ${\cal E}=\{p_s,\rho_s\}$,
namely, it emits a system in a quantum state $\rho_s$ with 
probability $p_s>0$. One of the fundamental questions in 
quantum information theory is 
to identify the optimal compression rate of ${\cal E}$,
namely, to determine
how much qubits are needed to 
compress a sequence of systems independently
prepared from this source so
that it can be decompressed back with negligible
errors in the asymptotic limit of the infinitely long sequence. 
Noting
that 
 the original states $\rho_s$ are reproduced 
after the decompression,
 the present results can be applied to 
the whole operation of compression and decompression,
and reveals the optimal compression rate in the blind scenario.
Using the decomposition 
$\rho_s
=\bigoplus_l {p}^{(s,l)}\rho^{(s,l)}_{\rm J} \otimes
\rho^{(l)}_{\rm K}$, the average density operator
$\rho\equiv\sum_s p_s\rho_s$ is also
decomposed as
\begin{equation}
\rho=\bigoplus_l p^{(l)} \rho^{(l)}_{\rm J}\otimes \rho^{(l)}_{\rm K},
\label{totalrhodec}
\end{equation}
where $p^{(l)}\equiv\sum_s p_s p^{(s,l)}$ and 
$\rho^{(l)}_{\rm J}\equiv (\sum_s p_s p^{(s,l)}
\rho^{(s,l)}_{\rm J})/p^{(l)}$.
This naturally gives a decomposition of the von Neumann entropy
of $\rho$, defined as $S(\rho)\equiv -\mbox{Tr}\rho \log_2 \rho$,
 into the sum of three parts as follows,
\begin{eqnarray}
S(\rho)&=&
\sum_l p^{(l)}\left[
-\log_2 p^{(l)}+S(\rho^{(l)}_{\rm J})+
S(\rho^{(l)}_{\rm K})
\right]
\nonumber \\
&\equiv& I_{\rm C}+I_{\rm NC}+I_{\rm R}.
\label{vondec}
\end{eqnarray}
Then, 
the form $\rho_s
=\bigoplus_l {p}^{(s,l)}\rho^{(s,l)}_{\rm J} \otimes
\rho^{(l)}_{\rm K}$ tells us that 
${\cal E}$ can be compressed into $I_{\rm C}+I_{\rm NC}$
qubits, and its optimality can be shown from  
the fact that any compression-decompression
scheme must be written in the form 
$U_{\rm AE}=
\bigoplus_l\bbox{1}^{(l)}_{\rm J}\otimes U^{(l)}_{\rm
KE}$ \cite{compressibility}.
It was also shown  that among $I_{\rm C}+I_{\rm NC}$
qubits, $I_{\rm C}$ qubits can be replaced by 
the same number of classical bits \cite{jozsa,cost}.
A similar argument can also be made to the teleportation 
of the ensemble ${\cal E}$, and the optimally required amount 
of entanglement was shown to be $I_{\rm NC}$ ebits.
These results again suggest that the decomposition 
$\rho_s
=\bigoplus_l {p}^{(s,l)}\rho^{(s,l)}_{\rm J} \otimes
\rho^{(l)}_{\rm K}$ gives a way to classify the 
degrees of freedom into the three parts, namely, 
 classical, 
nonclassical, and redundant parts.

Using 
Theorem 4 derived in Sec.~\ref{sec:properties}, 
we immediately see that the above information-theoretic 
functions  $I_{\rm C}({\cal E})$,
$I_{\rm NC}({\cal E})$, $I_{\rm R}({\cal E})$,
and hence the various optimal rates, 
are additive for independent sources. That is 
to say, if we consider another source 
${\cal E}'=\{q_{s'},\sigma_{s'}\}$ and the combined 
source $\tilde{\cal E}=\{p_s q_{s'},\rho_s\otimes \sigma_{s'}\}$,
we have $I_X(\tilde{\cal E})=
I_X({\cal E})+I_X({\cal E}')$ $(X={\rm C}, {\rm NC},{\rm R})$.

\section{Conclusion}

In this paper, we have considered a situation which we frequently
encounter in dealing with problems in quantum information, namely,
given a system secretly prepared in one of the possible states
$\{\rho_s\}$, conducting a general operation to the system, then leaving
the state of the system exactly in the same state as the initially given 
state. In order to derive a general property of such operations,
we noted two basic principles: One is a natural extension of a property 
of classical signals, which states that in order not to disturb 
a signal which may be produced by two different probability distributions,
 we are not allowed to operate on the entire signal space freely,
but are forced to operate on two or more signal subspaces independently.
The other principle stems genuinely from quantum origin, and it states 
that if we are to operate on two subspaces independently while 
preserving a state having a nonzero off-diagonal part with respect to 
the two subspaces, the operations to the two subspaces must satisfy 
a similarity defined through the off-diagonal part of the state.
The two types of constraints alternately invokes each other, 
and finally reveals a stringent condition for the operations 
to preserve $\{\rho_s\}$, together with a decomposition of 
the support space of $\{\rho_s\}$, which takes a form 
${\cal H}_{\rm A}=
\bigoplus_l{\cal H}^{(l)}_{\rm J}\otimes {\cal H}^{(l)}_{\rm
K}$. Under this decomposition, the states $\{\rho_s\}$
are written as $\rho_s
=\bigoplus_l {p}^{(s,l)}\rho^{(s,l)}_{\rm J} \otimes
\rho^{(l)}_{\rm K}$.
If we consider how 
 the information of the state index
$s$ is encoded on three parts, namely, on index $l$, 
on Hilbert space ${\cal H}^{(l)}_{\rm J}$, and 
on Hilbert space ${\cal H}^{(l)}_{\rm K}$,
we may regard them as 
  classical, nonclassical (quantum),
and redundant parts, respectively, since $\rho_s$ has no off-diagonal part
with respect to index $l$, and no information on $s$ is stored on ${\cal
H}^{(l)}_{\rm K}$. Under this decomposition,
the main result describing the property of the operations 
to preserve $\{\rho_s\}$ is written as 
$U_{\rm AE}=
\bigoplus_l\bbox{1}^{(l)}_{\rm J}\otimes U^{(l)}_{\rm
KE}$, which informally implies that 
the nonclassical part is untouchable, the classical part 
is read-only, and the redundant part is open. 

The result may be viewed as an unexpectedly straightforward 
extension of the simplest case of binary string $s=0,1$ encoded on
 two pure states $|\Psi_0\rangle$ and $|\Psi_1\rangle$. We can 
distinguish three cases according to the inner product of the two 
pure states.
The encoding will be regarded as `classical' when the two states are
orthogonal, `nonclassical' when they are nonorthogonal and nonidentical,
and `redundant' when identical. When this situation is extended to 
allow mixed states and a larger number of states, it has turned out
that the three types of the encoding may coexist, but they are still
distinct. The inner product for two vectors must be replaced 
by mathematical concepts describing rather complicated 
 relations among
many density operators. In this paper, we have attempted to do this by
regarding  the Hilbert space
as a module  over an algebra generate by $\{\rho_s\}$ with a proper
normalization. Then, the notion of `nonorthogonal' corresponds to 
irreducibility (being simple) of a submodule, the notion of
`orthogonal' corresponds to reducibility into inequivalent simple 
submodules, and the notion of `identical' corresponds to reducibility 
into equivalent simple submodules.
 
The main result was shown to be applicable to various problems 
of quantum information. The tasks of cloning, broadcasting, imprinting,
and eavesdropping in quantum cryptography belong to a class of 
problems in which extraction of information on the initial 
state of the system is required without introducing disturbance. 
The present result can naturally be applied to this class of 
problems, and helps to derive various conditions on the set of possible 
initial states for various tasks to be feasible.
In addition, the result was also successfully applied to 
tasks such as quantum data compression and quantum teleportation, in
which  the extraction of the information on the initial 
state is not directly required. It was shown 
(see also \cite{compressibility,cost}) that the optimal rates  
of bits and qubits for asymptotically faithful blind compression 
is simply equal to the Shannon or von Neumann entropy of 
the classical and nonclassical parts, respectively. 
This result also justifies the terminology of classical,
nonclassical, and redundant parts in operational sense,
namely, the classical part can be encoded on bits and 
sent through a classical channel, but the nonclassical part 
can be encoded only on qubits and requires shared entanglement 
to be sent over a classical channel.

\section*{Acknowledgment}

This work was supported by a Grant-in-Aid for Encouragement of Young
Scientists (Grant No.~12740243) and a Grant-in-Aid for Scientific 
Research (B) (Grant No.~12440111)
by Japan Society for the Promotion of
Science.

\end{document}